\documentclass[a4paper,fleqn,usenatbib]{mnras}

\usepackage{graphicx, color}
\usepackage{hyperref}
\usepackage{epsfig}
\usepackage{rotate}
\usepackage{latexsym}
\usepackage{amssymb}
\usepackage{amsmath}
\usepackage{psfrag}
\usepackage{xspace}
\newcommand{\nn}{\nonumber}

\newcommand{\be}{\begin{equation}}
\newcommand{\ee}{\end{equation}}
\newcommand{\bdm}{\begin{displaymath}}
\newcommand{\edm}{\end{displaymath}}
\newcommand{\bea}{\begin{eqnarray}}
\newcommand{\eea}{\end{eqnarray}}
\newcommand{\bt}{\begin{tabular}}
\newcommand{\et}{\end{tabular}}
\newcommand{\lan}{\langle}
\newcommand{\ran}{\rangle}
\newcommand{\kv}{{\bf k}}

\newcommand{\qv}{{\bf q}}

\newcommand{\kmax}{k_{\rm max}}

\def\d{\delta}
\def\Mpc{\, h^{-1} \, {\rm Mpc}}
\def\kMpc{\, h \, {\rm Mpc}^{-1}}
\def\Ms{\, h^{-1} \, {\rm M}_{\odot}}

\def\fun#1#2{\lower3.6pt\vbox{\baselineskip0pt\lineskip.9pt
        \ialign{$\mathsurround=0pt#1\hfill##\hfil$\crcr#2\crcr\sim\crcr}}}
\def\la{\mathrel{\mathpalette\fun <}}
\def\ga{\mathrel{\mathpalette\fun >}}

\newcommand{\eq}[1]{eq.~(\ref{#1})}
\newcommand{\ie}{{\it i.e.}}
\newcommand{\eg}{{\it e.g.}\xspace}

\def\halogen{{\sc{Halogen}}\xspace}
\def\cola{{\sc{ICE-COLA}}\xspace}
\def\patchy{{\sc{Patchy}}\xspace}
\def\pin{{\sc{Pinocchio}}\xspace}
\def\peak{{\sc{PeakPatch}}\xspace}

\title[Comparing approximate methods]{
Comparing approximate methods for mock catalogues and covariance matrices III: Bispectrum }
\author[M. Colavincenzo et al.]{Manuel Colavincenzo$^{1,2,3}$\thanks{E-mail:mcolavin@to.infn.it},
Emiliano Sefusatti$^{4,5}$,
Pierluigi Monaco$^{3,4,5}$,
\newauthor
Linda Blot$^{6,7}$,
Martin Crocce$^{6,7}$,
Martha Lippich$^{8,9}$,
Ariel G. S\'{a}nchez$^{8}$,
\newauthor
Marcelo A. Alvarez$^{10}$,
Aniket Agrawal$^{11}$,
Santiago Avila$^{12}$,
\newauthor
Andr\'es Balaguera-Antol\'{i}nez$^{13,14}$,
Richard Bond$^{15}$,
Sandrine Codis$^{15,16}$,
\newauthor
Claudio Dalla Vecchia$^{13,14}$,
Antonio Dorta$^{13,14}$,
Pablo Fosalba$^{6,7}$,
\newauthor
Albert Izard$^{17,18,6,7}$,
Francisco-Shu Kitaura$^{13,14}$,
Marcos Pellejero-Ibanez$^{13,14}$,
\newauthor
George Stein$^{15}$,
Mohammadjavad Vakili$^{19}$,
Gustavo Yepes$^{20,21}$
\\
\vspace{0.1cm}\\
$^{1}$ Dipartimento  di  Fisica,  Universit\`a  di  Torino,  Via  P.  Giuria  1,  10125  Torino,  Italy\\
$^{2}$ Istituto  Nazionale  di  Fisica  Nucleare,  Sezione  di  Torino,  Via  P.  Giuria  1,  10125  Torino,  Italy\\
$^{3}$ Dipartimento di Fisica, Sezione di Astronomia, Universit\`a di Trieste, via Tiepolo 11, 34143 Trieste, Italy\\
$^{4}$ Istituto  Nazionale  di  Astrofisica, Osservatorio Astronomico di Trieste, via Tiepolo 11, 34143 Trieste, Italy\\
$^{5}$ Istituto  Nazionale  di  Fisica  Nucleare, Sezione di Trieste, Via Valerio, 2, 34127 Trieste, Italy\\
$^{6}$ Institute of Space Sciences (ICE, CSIC), Campus UAB, Carrer de Can Magrans, s/n,  08193 Barcelona, Spain \\
$^{7}$ Institut d'Estudis Espacials de Catalunya (IEEC), 08034 Barcelona, Spain \\
$^{8}$ Max-Planck-Institut f\"ur extraterrestrische Physik, Postfach 1312, Giessenbachstr., 85741 Garching, Germany\\
$^{9}$ Universit\"ats-Sternwarte M\"unchen, Ludwig-Maximilians-Universit\"at M\"unchen, Scheinerstrasse 1, 81679 Munich, Germany\\
$^{10}$ Berkeley Center for Cosmological Physics, Campbell Hall 341, University of California, Berkeley CA 94720, USA \\
$^{11}$ Max-Planck-Institut f\"ur Astrophysik, Karl-Schwarzschild-Str. 1, 85741 Garching, Germany\\
$^{12}$ Institute of Cosmology \& Gravitation, Dennis Sciama Building, University of Portsmouth, Portsmouth PO1 3FX, UK\\
$^{13}$ Instituto de Astrof\'isica de Canarias, C/V\'ia  L\'actea, s/n, 38200, La Laguna, Tenerife, Spain\\
$^{14}$ Departamento Astrof\'isica, Universidad de La Laguna,  38206 La Laguna, Tenerife, Spain\\
$^{15}$ Canadian Institute for Theoretical Astrophysics, University of Toronto, 60 St. George Street, Toronto, ON M5S 3H8, Canada\\
$^{16}$ Institut d'Astrophysique de Paris, CNRS \& Sorbonne Universit\'e, UMR 7095, 98 bis boulevard Arago, 75014 Paris, France\\
$^{17}$ Jet Propulsion Laboratory, California Institute of Technology, 4800 Oak Grove Drive, Pasadena, CA 91109, USA\\
$^{18}$ Department of Physics and Astronomy, University of California, Riverside, CA 92521, USA\\
$^{19}$ Leiden Observatory, Leiden University, P.O. Box 9513, NL-2300 RA, Leiden, The Netherlands \\
$^{20}$ Departamento de F\'isica Te\'orica, M\'odulo 15, Universidad Aut\'onoma de Madrid, 28049 Madrid, Spain\\
$^{21}$ Centro de Investigaci\'on Avanzada en F\'isica Fundamental (CIAFF), Universidad Aut\'onoma de Madrid, 28049 Madrid, Spain\\\vspace{-0.1cm}\\
$^\star${\tt E-mail:mcolavin@to.infn.it}}
\date{Accepted XXX. Received YYY; in original form ZZZ}
\pubyear{2018}
\begin{document}
\maketitle
\label{firstpage}

\begin{abstract}
We compare the measurements of the bispectrum and the estimate of its
covariance obtained from a set of different methods for the efficient
generation of approximate dark matter halo catalogs to the same
quantities obtained from full N-body simulations. To this purpose we
employ a large set of three-hundred realisations of the same cosmology
for each method, run with matching initial conditions in order to
reduce the contribution of cosmic variance to the comparison. In
addition, we compare how the error on cosmological parameters such as
linear and nonlinear bias parameters depends on the approximate method
used for the determination of the bispectrum variance. As general result, most methods provide errors within 10\% of the errors estimated from N-body simulations. Exceptions are those methods requiring calibration of the clustering amplitude but restrict this to two-point statistics.  Finally we test how our results are affected by being limited to a few hundreds
measurements from N-body simulation by comparing with a larger set of several thousands
realisations performed with one approximate method.
\end{abstract}


%%%%%%%%%%%%%%%%%%%%%%%%%%%%%%%%%%%%%%%%%%%%%%%%%%%%%%%
%%%%%%%%%%%%%%%%%%%%%%%%%%%%%%%%%%%%%%%%%%%%%%%%%%%%%%%
\section{Introduction}

This is the last of a series of three papers exploring the problem of
covariance estimation for large-scale structure observables based on
dark matter halo catalogs obtained from approximate methods. The
importance of a large-set of galaxy catalogs both for purposes of
covariance estimation as well as for the testing of the analysis
pipeline has become evident over the last decade when such tools have
been routinely employed in the exploitation of several major galaxy
surveys \citep[see, \eg][]{ManeraEtal2013, DelaTorreEtal2013,
  KodaEtal2016, KitauraEtal2016, AvilaEtal2017a}.

In this context, it is crucial to ensure that mock catalogs correctly
reproduce the statistical properties of the galaxy distribution. Such
properties are characterised not only by the two-point correlation
function, but are quantified as well in terms of higher-order
correlators like the 3-point and 4-point correlation functions, since
the large-scale distributions of both matter and galaxies are highly
non-Gaussian random fields.

A correct non-Gaussian component in mock galaxy catalogs has
essentially two important implications. In the first place, we expect
the trispectrum, {\ie} the 4-point correlation function in Fourier
space, to contribute non-negligibly to the covariance of two-point
statistics. This is perhaps more evident in the case of the power
spectrum, already in terms of the direct correlation between band
power that we measure even in the ideal case of periodic box
simulations \citep[see, \eg][]{MeiksinWhite1999,
  ScoccimarroZaldarriagaHui1999, TakahashiEtal2009, NganEtal2012,
  BlotEtal2015,ChanBlot2017}. In addition, finite-volume effects such
as beat-coupling/super-sample covariance
\citep{HamiltonRimesScoccimarro2006, RimesHamilton2006,
  SefusattiEtal2006, TakadaHu2013} and local average of the density
field \citep{DePutterEtal2012} can be described as consequences of the
interplay between the survey window function and both the galaxy
bispectrum and trispectrum. In the second place higher-order
correlation functions, and particularly the galaxy 3PCF and the
bispectrum are emerging as relevant observables in their own right,
capable of complementing the more standard analysis of 2PCF and power
spectrum \citep{GaztanagaEtal2009, SlepianEtal2017, GilMarinEtal2015,
  GilMarinEtal2015B, GilMarinEtal2017,
  PearsonSamushia2017,2017arXiv170902473C}.

Both these aspects provide strong motivations for ensuring that not
only higher-order correlations are properly reproduced in mock catalogs
but also their own covariance properties are recovered with sufficient
accuracy. In this work we focus, in particular, on the bispectrum of
the halo distribution. This is the lowest order non-Gaussian statistic
characterising the three-dimensional nature of the large-scale
structure. It has also the practical advantage of requiring relatively
small numerical resources for its estimation on large sets of
catalogs, at least with respect to the 3-point correlation function in
real space. On the other hand, a correct prediction of the halo
bispectrum does {\em not} ensure that higher-order correlators such as
the halo trispectrum are similarly accurately reproduced. For
instance, a matter distribution realised at second order in Lagrangian
Perturbation Theory (LPT, the basis for several approximate methods)
is characterised by a bispectrum fully reproducing the expected
prediction at tree-level in Eulerian PT valid at large scales but that
is not the case for the matter trispectrum since the scheme only
partially reproduces the third order Eulerian nonlinear correction \citep{Scoccimarro1998}.

With this caveat in mind, in this paper we focus on the direct
comparison of the halo bispectrum and its covariance, along with a
comparison of the errors on the recovered halo bias parameters from a
simple likelihood analysis adopting different estimates of the
bispectrum variance and covariance. Clearly our sets of 300 halo catalogs from N-body
simulations and the various approximate methods limits a proper
comparison at the covariance level, since a reliable estimate of the
covariance matrix requires thousands of such
realisations. Nevertheless we explore the implications of such
limitation taking advantage of a much larger set of 10,000 runs, used
for the first time in \citep{ColavincenzoEtal2017}, of one of the
approximate methods.

Two companion papers focus on similar comparisons for the 2-point
correlation function \citep{LippichEtal2018} and for the power
spectrum \citep{BlotEtal2018}: we will refer to them, respectively, as
Paper I and Paper II throughout this work.

This paper is organised as follows. In section~\ref{sec:catalogs} we
present the approximate methods considered in this work and how they
address the proper prediction of the non-Gaussian properties of the
halo distribution. In section~\ref{sec:bispm} we describe the
measurements of the halo bispectrum and its covariance for each set of
catalogs which are then compared in section~\ref{sec:comparison}. In
section~\ref{sec:comperrors} we extend the comparison to the errors on
cosmological parameters while in section~\ref{sec:tests} we
present a few tests to quantify possible systematics due to the
limited number of catalogs at our disposal. Finally, we present our
conclusions in section~\ref{sec:conclusions}.

%%%%%%%%%%%%%%%%%%%%%%%%%%%%%%%%%%%%%%%%%%%%%%%%%%%%%%%
%%%%%%%%%%%%%%%%%%%%%%%%%%%%%%%%%%%%%%%%%%%%%%%%%%%%%%%
\section{The catalogs}
\label{sec:catalogs}

For a detailed description of the different approximate methods
compared in this, as well as the two companion papers, we refer the reader to
section 3 of Paper I, while for a more general examination of the
state-of-the-art in the field we refer to the review in
\citet{Monaco2016}. For a quick reference we reproduce in
Table~\ref{tab:mets} the Table 1 of Paper II, providing a brief
summary of the codes considered. Here we briefly discuss the main
characteristics of the catalogs and the implications for accurate
bispectrum predictions.\\

\begin{table*}
\begin{center}
\begin{tabular}{l l l l}
\hline
Method   &  Algorithm        &    Computational Requirements   &  Reference  \\ 
\hline   
Minerva     & {\bf N-body}        &  CPU Time: $4500$ hours                 &\cite{GriebEtal2016}  \\
         & Gadget-2  			&  Memory allocation: $660$ Gb
         & {\it https://wwwmpa.mpa-garching.mpg.de/} \\
         & Halos  : SubFind         & 		& {\it gadget/} \\ 
\hline
\hline   
\cola     & Predictive        &  CPU Time:    $33$ hours              &\cite{IzardCrocceFosalba2016}  \\
         & 2LPT + PM solver  	&  Memory allocation: $340$ Gb   &
         Modified version of:  \\
         & Halos  : FoF(0.2)         &
         & {\it https://github.com/junkoda/cola\_halo}\\
\hline
\pin     & Predictive        &  CPU Time:  $6.4$ hours                &\cite{MonacoEtal2013, MunariEtal2017}  \\
         & 3LPT + ellipsoidal collapse			  &  Memory allocation:  $265$ Gb
         & {\it https://github.com/pigimonaco/Pinocchio} \\
         & Halos  :  ellipsoidal collapse        &              & \\
\hline
\peak     & Predictive        &  CPU Time:  $1.72$ hours$^*$                &\cite{BondMyers1996A, BondMyers1996B, BondMyers1996C}  \\
         & 2LPT + ellipsoidal collapse 			  &  Memory allocation:   $75$ Gb$^*$          & {\rm Not public}  \\
         & Halos  : Spherical patches 
         &             & \\
         & over initial overdensities         &           & \\
\hline
\halogen     & Calibrated        &  CPU Time:  $0.6$ hours & \cite{AvilaEtal2015}.  \\
         & 2LPT + biasing scheme			  &  Memory allocation: $44$ Gb            & {\it https://github.com/savila/halogen}  \\
         & Halos  :  exponential bias        &   Input: $\bar{n}$, 2-pt correlation function    & \\
         &	& halo masses and velocity field & \\
\hline
\patchy     & Calibrated        &  CPU Time:  $0.2$ hours                &\cite{KitauraYepesPrada2014}  \\
         & ALPT + biasing scheme			  &  Memory allocation:  $15$
         Gb           & {\rm Not Public}  \\
         & Halos  : non-linear, stochastic        &    Input: $\bar{n}$, halo masses and       & \\
         & and scale-dependent bias 	& environment \citet{ZhaoKitauraEtal2015}		& \\
\hline
Lognormal     & Calibrated        &  CPU Time:  $0.1$ hours                &\cite{AgrawalEtal2017}  \\
         & Lognormal density field 			  &  Memory
         allocation:  $5.6$ Gb           & {\it
           https://bitbucket.org/komatsu5147/}  \\       
         & Halos  :  Poisson sampled points        				&      Input:  $\bar{n}$, 2-pt correlation function     & {\it
           lognormal\_galaxies}  \\
\hline
Gaussian     & Theoretical       &  CPU Time:  n/a                & \citet{ScoccimarroEtal1998} for the bispectrum \\
         & Gaussian density field 			  &  Memory allocation:  n/a           &  \\
         & Halos  :  n/a        				&     Input:   $P(k)$ and $\bar{n}$    &  \\
\hline
\end{tabular}
\end{center}
\label{tab:mets}
\caption{Name of the methods, type of algorithm, halo definition,
  computing requirements and references for the compared methods. All
  computing times are given in cpu-hours per run and memory
  requirements are per run, not including the generation of the
  initial conditions. The computational resources for halo finding in
  the N-body and \cola mocks are included in the requirements. The
  computing time refers to runs down to redshift 1 except for the
  N-body where we report the time down to redshift 0 (we estimate an
  overhead of $\sim$50$\%$ between $z=0$ and $z=1$. Since every code
  was run in a different machine the computing times reported here are
  only indicative. We include the information needed for
  calibration/prediction of the covariance where relevant. Mocks
  marked with ``$^*$'' require an higher resolution run in order to
  resolve the lower mass halos of our Sample 1 and therefore more
  computational resources than quoted here.}
\end{table*}

For all runs we consider a box of size $L=1500\Mpc$ and a cosmology
defined by the best-fitting parameters of the analysis in
\citet{SanchezEtal2013}. The N-body runs employ a number of particles
of 1000$^3$ leading to a particle mass
$m_p=2.67\times10^{11}\Ms$. In addition to the 100 runs mentioned in \citet{GriebEtal2016}, for this work we consider additional simulations for a total of 300 runs. 

We work on the halo catalogs obtained from the N-body identified with
a standard Friends-of-friends (FoF) algorithm. FoF halos were then
subject to the unbinding procedure provided by the Subfind algorithm
\citep{SpringelEtal2001} from snapshots at $z = 1$. We consider two
samples characterised by a minimal mass of $M_{min}=42
\,m_p=1.12\times 10^{13}\Ms$ (Sample 1) and $M_{min}=100
\,m_p=2.67\times 10^{13}\Ms$ (Sample 2). The corresponding number
densities are of $2.13 \times 10^{-4}$ and $5.44\times 10^{-5}$,
respectively. For Sample 2 the power spectrum signal is dominated by
shot-noise for scales $k\gtrsim 0.15\kMpc$, while for Sample 1 the shot-noise
contribution is always below the signal but still not negligible.

We produced a set of 300 realisations with each of the approximate
methods considered, imposing the same initial conditions as the N-body
runs in order to reduce any difference due to cosmic variance. The
definition of the two samples in the catalogs obtained by the
approximate methods depends on the specific algorithm.

We can distinguish between three different classes of algorithms:
\emph{predictive, calibrated} and \emph{analytical
  methods}. Predictive methods (\cola, \pin and \peak) aim at
identifying the Lagrangian patches that collapse into halos and do not
need to be recalibrated against a simulation. In particular,
  \cola is a PM solver, so it is expected to be more accurate at a
  higher computational cost \citep[see][]{IzardCrocceFosalba2016} . We
  choose a number of steps that sets its numerical requirements in
  between those of \pin run and of a full N-body simulation. Calibrated methods
(\halogen and \patchy) populate a large-scale density field with halos
using a bias model, and need to be recalibrated to match a sample in
number density and clustering amplitude. We should remark that
  while \halogen is calibrated only at the level of two-point
  statistics while \patchy extends its calibration to the three-point
  function in configuration space \citep{VakiliEtal2017}. In addition,
  all calibrations are performed in real, not redshift,
  space. Analytical methods include the Gaussian prediction for the
bispectrum covariance based on the measured power spectrum, and the
Lognormal method, predicting the halo distribution from some
assumption on the density field PDF. In particular, the Lognormal
realisations do not share the same initial conditions as the N-body
runs. Therefore, we employ for this method, the covariance estimated
from 1,000 realisation in order to beat down sample variance.

Notice that also for the predictive methods the minimal mass for each
sample is set by requiring the same abundance as the N-body samples. A
comparison that assumes directly the same mass thresholds as the
N-body samples is discussed in appendix~\ref{app:masscuts}. All other
methods assume such density matching by default.  For the \peak
comparison, only the larger mass sample is available.

All methods, with the exception of Lognormal, employ Lagrangian PT at
second order (or higher) to determine the large-scale matter density
field. We expect therefore, as mentioned before, that at least at
large scales where the characteristic LPT suppression of power is
still negligible, the measured halo bispectrum presents qualitatively
the expected dependence on the shape of the triangular
configurations. Any difference with the full N-body results at large
scales will likely arise from the specific way each method implements
the relation between 2LPT-displaced matter particles and its
definition of halos or particle groups. The case of Lognormal is
different since it is based on a nonlinear transformation of the
Gaussian matter density qualitatively reproducing the nonlinear
density probability distribution function \citep{ColesJones1991}, but
with no guarantee to properly reproduce the proper dependence on
configuration of higher-order correlation functions, starting from the
matter bispectrum.

These considerations have been already illustrated by the results of
the code-comparison project of \citet{ChuangEtal2015}. This work
comprises a comparison of both the 3-Point Correlation Function and
the bispectrum of halos of minimal mass of $10^{13}\Ms$, very similar
to one of the two samples considered in our work, but evaluated at the
lower redshift $z\simeq 0.55$. Each measurement was performed for a
relatively small set of configurations, covering, in the bispectrum
case, the range of scales $0.1\le (k/ \kMpc)\le 0.3$. The codes \cola,
{\sc{EZmock}} \citep{ChuangEtal2015A} and \patchy (the last two requiring
calibration of the halo power spectrum) reproduced the N-body results
with an accuracy of 10-15\%, \pin at the 20-25\% level, while
\halogen and {\sc{PThalos}} \citep{ScoccimarroSheth2002, ManeraEtal2013} at
the 40-50\%. All these methods correctly recovered the overall shape
dependence. On the other hand, the Lognormal method failed to do so,
despite the predicted bispectrum showed a comparable, overall
magnitude \citep[see also][]{WhiteTinkerMcBride2014}. It should be
noted that in some cases, as \eg ~\pin, the codes employed in
this work correspond to an updated version w.r.t. those considered by
\citet{ChuangEtal2015}. 

We notice that, in the present work, we will go beyond the results of \citet{ChuangEtal2015}, extending the test of the approximate methods to the comparison of the recovered bispectrum {\em covariance}.

%%%%%%%%%%%%%%%%%%%%%%%%%%%%%%%%%%%%%%%%%%%%%%%%%%
%%%%%%%%%%%%%%%%%%%%%%%%%%%%%%%%%%%%%%%%%%%%%%%%%%
\section{Measurements}
\label{sec:bispm}

For each sample we estimate the Fourier-space density on a grid of 256 of linear size employing the 4th-order interpolation algorithm and the interlacing technique implemented in the {\sc{PowerI4}}\footnote{\url{https://github.com/sefusatti/PowerI4}} code described in \citet{SefusattiEtal2016}.    

The bispectrum estimator is given by
\bea
\hat{B}_{tot}(k_1,k_2,k_2) & \equiv & \frac{k_f^3}{V_B(k_1,k_2,k_3)}\int_{k_1}\!\! \!\!d^3q_1\!\!\int_{k_2} \!\!\!\!d^3q_2\!\!\int_{k_3} \!\!\!\!d^3q_3 \nn\\
& & \times \,\d_D(\qv_{123})\,\d_{\qv_1}\,\d_{\qv_2}\,\d_{\qv_3}
\eea
where the integrations are taken on shells of size $\Delta k$ centered on $k_i$ and where
\bea\label{eq:VB}
V_B(k_1,k_2,k_2) & \equiv & \int_{k_1}\!\! d^3q_1\int_{k_2} \!\!d^3q_2\int_{k_3} \!\!d^3q_3\,\d_D(\qv_{123}) \nn\\
& \simeq &  8\pi^2\ k_1 k_2 k_3 \Delta k^3
\eea
is a normalisation factor counting the number of fundamental triangles (those defined by the vectors $\qv_1$, $\qv_2$ and $\qv_3$ on the discrete Fourier density grid) in a given triangle bin (defined instead by the triplet $k_1$, $k_2$ and $k_3$ plus the size $\Delta k$ for all sides). Its implementation is based on the algorithm described in \citet{Scoccimarro2015}. 

The measured bispectrum will be affected by shot-noise. Under the assumption of Poisson shot-noise, we correct the measurement $\hat{B}$ as follows \citep{MatarreseVerdeHeavens1997}
\bea
\label{eq:Bshotnoise}
B(k_1,k_2,k_3) & = & \hat{B}_{tot} - \frac{1}{(2\pi)^3\bar{n}}[P(k_1) + P(k_2)+P(k_3)] \nn\\
& & - \frac{1}{(2\pi)^6\bar{n}^2} \,,
\eea
where $\bar{n}$ is the halo density of each individual catalog and $P(k)$ is the halo power spectrum, in turn corrected for shot-noise.

We consider all triangular configurations defined by discrete
wavenumbers multiples of $\Delta k=3 k_f$ with $k_f\equiv 2\pi/L$
being the fundamental frequency of the box, up to a maximum value of
$0.38\kMpc$, although we will limit our analysis to scales defined by
$k_i\le0.2\kMpc$, where we conservatively expect analytical
predictions in perturbation theory to accurately describe the galaxy
bispectrum. These choices lead to a total number of triangle bins of 508.
  
Given the estimator above, the Gaussian prediction for the variance is given by \citep{Scoccimarro2000B}, 
\bea
\Delta B^2 (k_1, k_2, k_3) & \equiv & \lan(\hat{B}^2-\lan\hat{B}\ran^2)\ran\nn\\
&  \simeq & s_B \frac{k^3_f}{V_B }  P_{tot}(k_1)P_{tot}(k_2) P_{tot}(k_3)\,,
\label{eqerror}
\eea
with $s_B = 6, 2, 1$ for equilateral, isosceles and scalene
triangles respectively and where $P_{tot}(k)=P(k)+1/[(2\pi)^3\bar{n}]$
includes the Poisson shot-noise contribution due to the halo density
$\bar{n}$. We will compare our measurements to this theoretical
prediction for the variance. For such comparison we will employ the
measured mean value of $P_{h,tot}(k)$ and the exact number of
fundamental triangles $V_B(k_1,k_2,k_3)$ as provided by the code,
which is slightly different, for certain triangular shapes, from the
approximate value on the second line of eq.~(\ref{eq:VB}).

Theoretical predictions are computed for ``effective''  values of the wavenumbers defined, for a given configuration of sides $k_1$, $k_2$ and $k_3$  by
\be
\tilde{k}_{1,23}\equiv \frac1{V_B}\int_{k_1}\!\! d^3q_1\,q_1\,\int_{k_2} \!\!d^3q_2\int_{k_3} \!\!d^3q_3\,\d_D(\qv_{123})\,,\ee
and similarly for the other two values. Differences with respect to evaluations at the center of each $k$-bin are marginally relevant and only so for the largest scales.

%%%%%%%%%%%%%%%%%%%%%%%%%%%%%%%%%%%%%%%%%%%%%%%%%%%%%%%
%%%%%%%%%%%%%%%%%%%%%%%%%%%%%%%%%%%%%%%%%%%%%%%%%%%%%%%
\section{Bispectrum and bispectrum covariance comparison}
\label{sec:comparison}

In this section we compare the measurements of the halo bispectrum for
the two halo samples both in real and redshift space. Since one of the
aims of this work is testing how accurately the non-Gaussian properties
of the large-scale halo distribution are recovered, it is relevant to
look at the lowest order non-Gaussian statistic also in real space,
while the bispectrum as a direct observable motivates all
redshift-space tests.

We compare as well the variance estimated from the 300 runs and the
covariance among different triangles. Clearly, 300 realisations are
not enough to provide a proper estimate of the covariance among 508
triplets. The comparison is then aiming at verifying that the same
statistical fluctuations appear across the estimates from different
approximate methods, taking advantage of the shared initial
conditions.

%%%%%%%%%%%%%%%%%%%%%%%%%%%%%%%%%%%%%%%%%%%%%%%%%%%%%%%%%%%%%%%%%%
\subsection{Real space}
\label{sec:comparison_real}

\begin{figure*}
\begin{center}
\includegraphics[width=0.95 \textwidth]{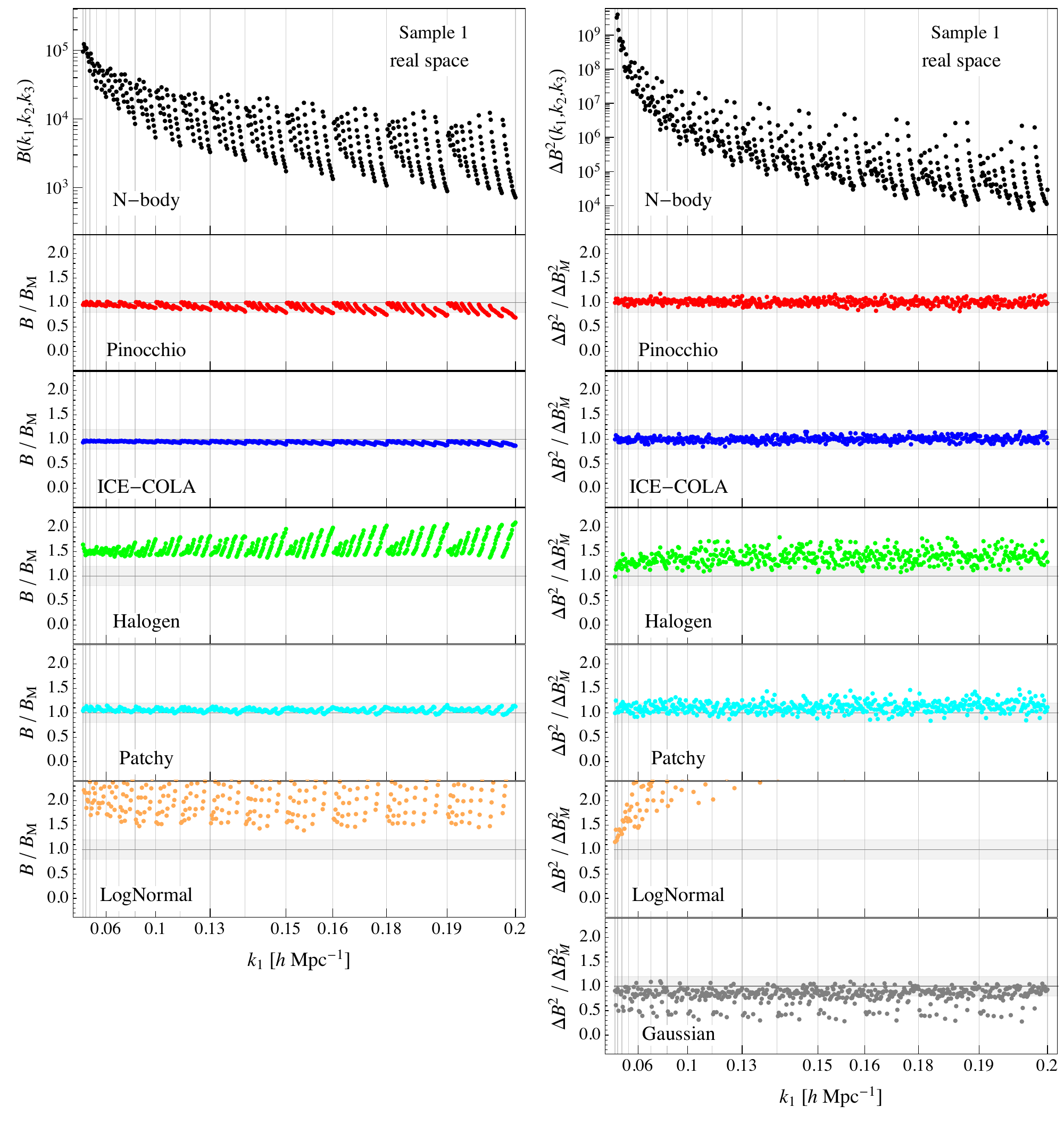}
\caption{\label{fig:bsRealAllD1} Average bispectrum (left column) and
  its variance (right column) for all triangle configurations obtained
  from the 300 realisations for the first mass sample in real
  space. The top panels show the results for the Minerva (black dots),
  while all other panels show the ratio between the estimate from an
  approximate method and the N-body one. In the last panel of the
  right column the grey dots show the ratio between the Gaussian
  prediction for the bispectrum variance, \eq{eqerror}, and the
  variance obtained from the N-body. The horizontal shaded area
  represents a 20$\%$ error. The vertical lines mark the triangle
  configurations where $k_1$ (the maximum of the triplet) is changing,
  so that all the points in between such lines correspond to all
  triangles with the same value for $k_1$ and all possible values of
  $k_2$ and $k_3$. Since we assume $k_1\ge k_2\ge k_3$, the value of
  $k_1$ corresponds also to the maximum side of the triangle.  Mocks
  for \peak are not provided in the first sample so its bispectrum is
  missing in this case.}
\end{center}
\end{figure*}

\begin{figure*}
\begin{center}
\includegraphics[width=0.95 \textwidth]{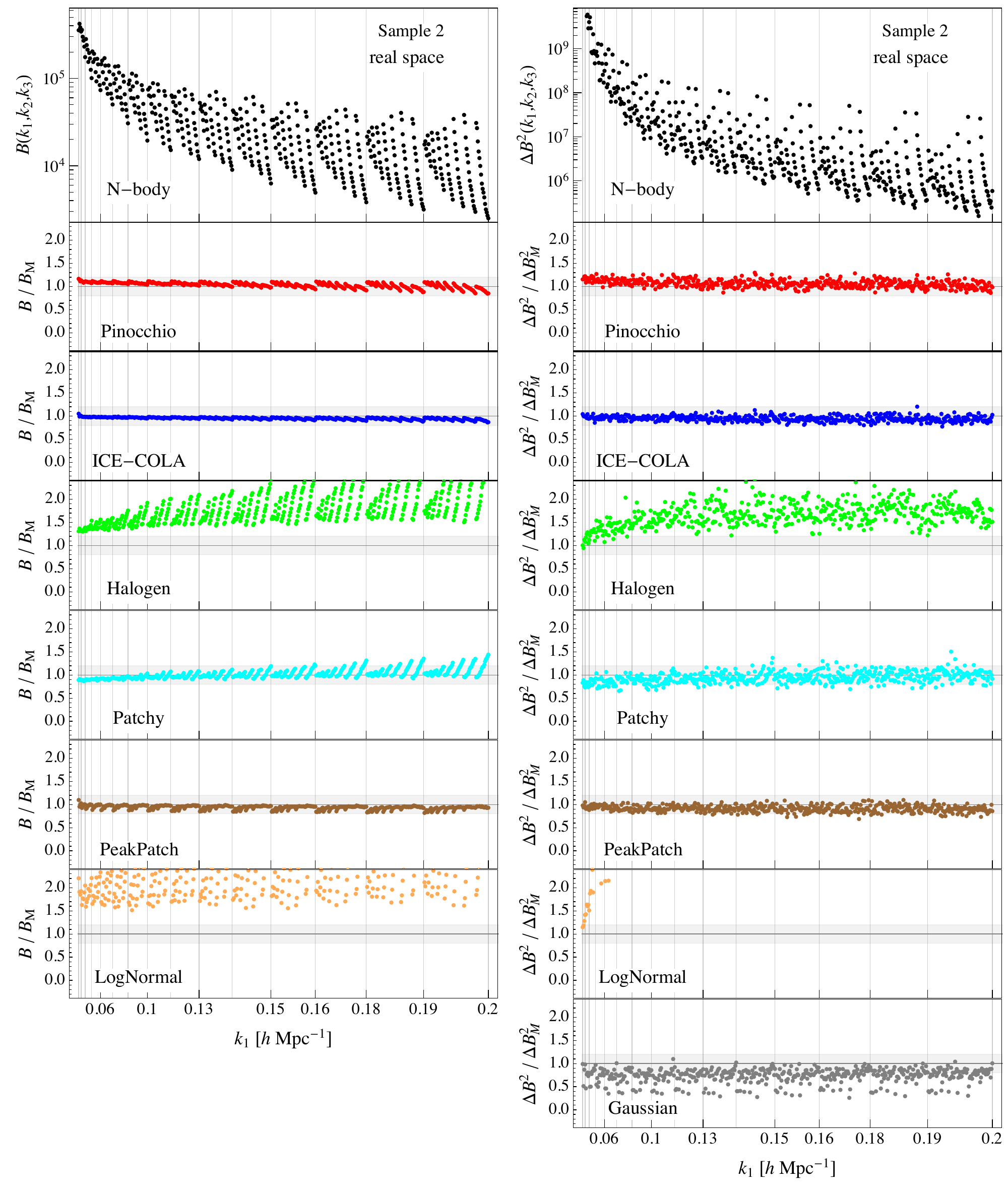}
\caption{\label{fig:bsRealAllD2} Same as figure~\ref{fig:bsRealAllD1}, but for Sample 2. }
\end{center}
\end{figure*}

Figures~\ref{fig:bsRealAllD1} and ~\ref{fig:bsRealAllD2} show,
respectively for Sample 1 and Sample 2, in the left column, top panel,
the real-space halo bispectrum averaged over the 300 N-body
simulations. The panels below show the ratio between the same
measurements obtained from all approximate methods and the N-body
results. The right column shows a similar comparison for the halo
bispectrum variance. For this quantity we include an additional,
bottom panel where we plot the comparison between the Gaussian
prediction for the bispectrum variance, eq. \ref{eqerror}, and the
N-body estimate. We will keep the color-coding for each methods
consistently throughout this paper.

Each dot represents the bispectrum for a particular triplet
$\left\{k_1, k_2,k_3\right\}$. These are plotted in an order where
$k_1\ge k_2\ge k_3$ with increasing values of each $k_i$ for all
allowed configurations. In practice, the first configurations are, in
units of the $k$-bin size $\Delta k$
\begin{displaymath}
\left\{1, 1,1\right\},~\left\{2, 1,1\right\},~\left\{2, 2,1\right\},~\left\{2, 2,2\right\},~\left\{3, 2,1\right\},~\dots
\end{displaymath}
The ticks on the abscissa mark the value of $k_1$, the largest wavenumber in each triplet, and the vertical grey lines denote the configurations where $k_1$ changes. 

All predictive methods, that is \pin, \cola and \peak (this last for
Sample 2), reproduce the N-body measurements within 15\% for most of
the triangle configurations, with some small dependence on the
triangle shape. Similar results, among the methods requiring some form
of calibration, are obtained for \patchy, with just some higher
discrepancies at the 20-30\% level appearing for Sample 2 at small
scales, mainly for nearly equilateral triangles. The other calibrated
methods fare worse. \halogen shows differences above 50\%, reaching
100\% for nearly equilateral configurations in both samples. The
LogNormal approach, as one can expect, shows the largest discrepancy
for all the scales and all the configurations in both samples.
 
Similar considerations can be made for the comparison of the
variance. In this case a large component is provided by the shot-noise
contribution, so the ratios to the N-body results show a less
prominent dependence on the triangle shape.  In general, we expect the
agreement with N-body to depend to a large extent, particularly for
Sample 2, on the correct matching of the object density, and more so
for those LPT-based methods that show a lack of power in this
regime. The Gaussian prediction underestimates the N-body result by
10-20$\%$ for the majority of configurations, and reaching up to 50\%
for squeezed triangles, \ie~those comprising the smallest wavenumber.

\begin{figure*}
\begin{center}
\includegraphics[width=0.95 \textwidth]{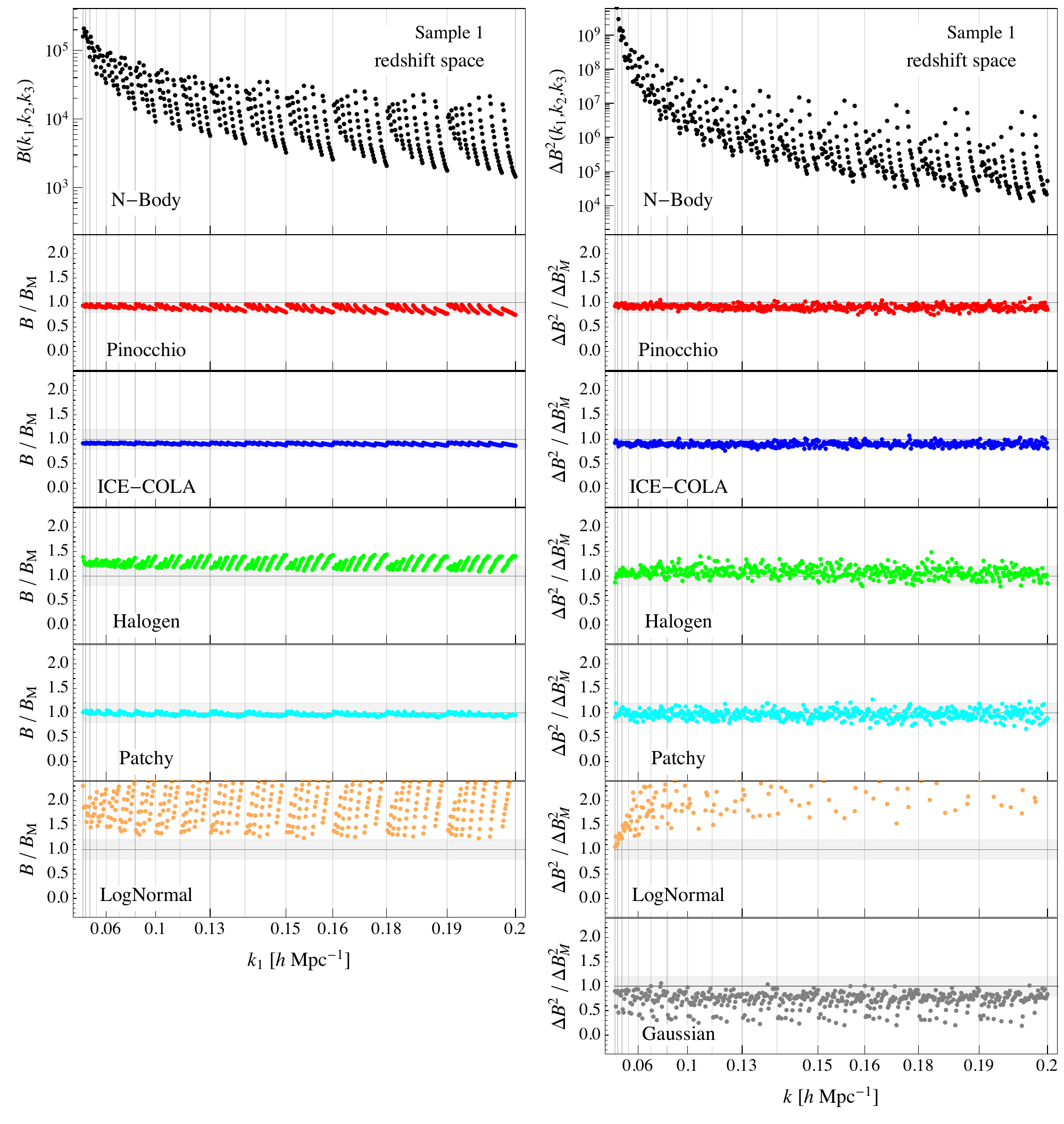}
\caption{\label{fig:bsRSAllD1} Average bispectrum (left column) and
  its variance (right column) for all triangle configurations obtained
  from the 300 realisations for the first mass sample in {\em
    redshift} space. The top panels show the results for the Minerva
  (black dots), while all other panels show the ratio between each a
  given estimate from an approximate method and the N-body one. In the
  last panel of the right column the grey dots show the ratio between
  the Gaussian prediction for the bispectrum variance, \eq{eqerror},
  and the variance obtained from the N-body. The horizontal shaded
  area represents a 20$\%$ error. The vertical lines mark the triangle
  configurations where $k_1$ (the maximum of the triplet) is
  changing. Mocks for \peak are not provided in the first sample
  so its bispectrum is missing in this case.}
\end{center}
\end{figure*}

\begin{figure*}
\begin{center}
\includegraphics[width=0.95 \textwidth]{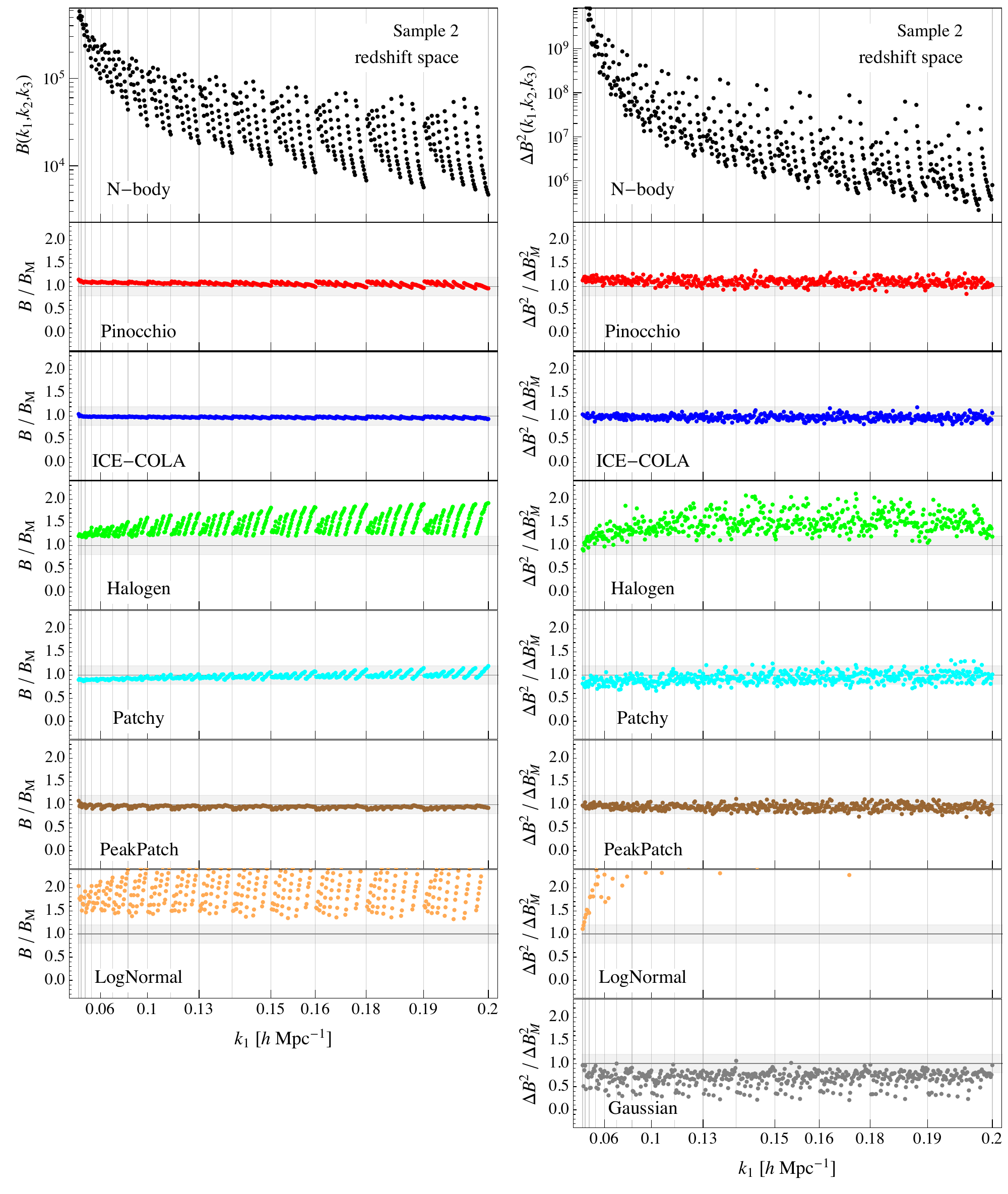}
\caption{\label{fig:bsRSAllD2} Same as figure~\ref{fig:bsRSAllD1}, but for Sample 2. }
\end{center}
\end{figure*}

\begin{figure*}
\begin{center}
\includegraphics[width=0.95\textwidth]{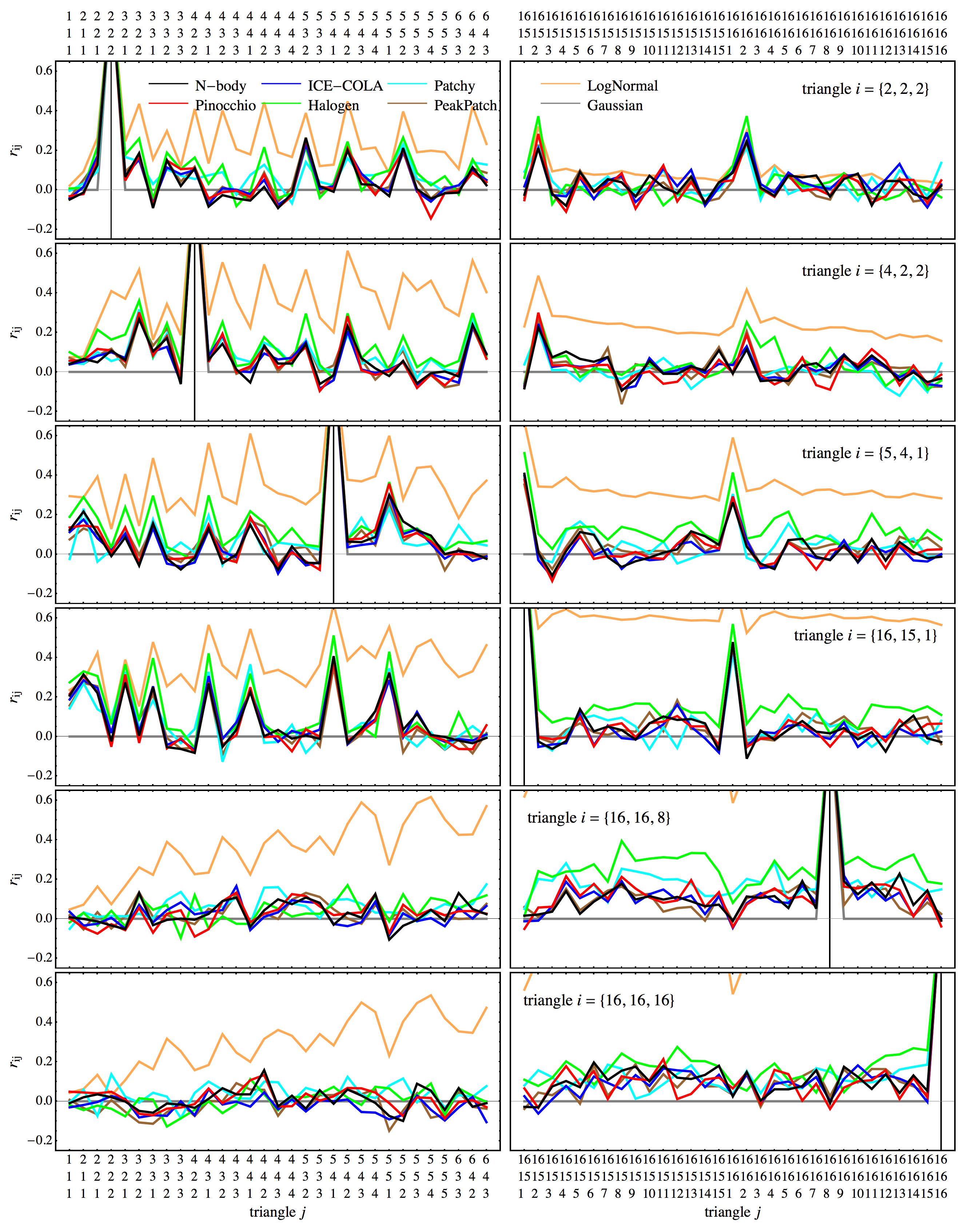}
\caption{\label{fig:bsRSccMrowsD1} Cross-correlation coefficients $r_{ij}$ for Sample 2, as defined in \eq{eqcross}, for a choice of six triangles $t_i$ (one for each row) against two subsets of configurations at large and small scales (left and right columns, respectively) in redshift space. See text for explanation.}
\end{center}
\end{figure*}

%%%%%%%%%%%%%%%%%%%%%%%%%%%%%%%%%%%%%%%%%%%%%%%%%%%%%%%%%%%%%%%%%
\subsection{Redshift space}

Figures \ref{fig:bsRSAllD1} and \ref{fig:bsRSAllD2}, respectively for the Sample 1 and 2, show the redshift-space bispectrum monopole (left column) and its variance (right column), with the same conventions assumed for the real-space results in figure
\ref{fig:bsRealAllD1}. The overall results are by and large very similar to the real-space ones. Only for the first sample, both \halogen and \patchy show a better agreement with the N-body results than in real space. As before Lognormal is the one that shows the largest disagreement with the N-body results.

Figure \ref{fig:bsRSccMrowsD1} shows, for Sample 1, a representative subset of the off-diagonal elements of the bispectrum covariance matrix in redshift space as estimated by the different methods. The quantities shown are the cross-correlation coefficients $r_{ij}$ defined as
\be
r_{ij} \equiv \frac{C_{ij}}{\sqrt{C_{ii}\,C_{jj}}}\,,
\label{eqcross}
\ee
where
\be
C_{ij}\equiv \lan(\hat{B}(t_i)-\lan\hat{B}(t_i)\ran)(\hat{B}(t_j)-\lan\hat{B}(t_j)\ran)\ran\,,
\ee
is the covariance between the bispectrum configuration $t_i=\left\{k_{i,1},k_{i,2},k_{i,3}\right\}$ and the configuration  $t_j=\left\{k_{j,1},k_{j,2},k_{j,3}\right\}$.

The figure shows the correlation of 6 chosen triangles $t_i$ with two
subsets of configurations $t_j$: one at large scale
$t_j=\left\{1,1,1\right\}\Delta k \dots~\left\{6,4,3\right\}\Delta k$
and one at small scales $t_j=\left\{16,15,1\right\}\Delta
k\dots~\left\{16,16,16\right\}\Delta k$, as explicitly denoted on the
abscissa in terms of triplets in units of $\Delta k$.

With the exception of the diagonal cases $t_i=t_j$, most of the
features in the $r_{ij}$ plots reflect random fluctuations rather
than actual correlations since 300 realisations are not sufficient to
accurately estimate the bispectrum covariance matrix. A more accurate
estimation of the matrix itself, limited to a single method, is
presented in section~\ref{sec:tests}, where we show how such
fluctuations are of the same order of the expected correlations among
triangles sharing, for instance, one or two sides, and it is therefore
impossible to tell them apart in this figure. Nevertheless, the random noise itself in the off-diagonal elements of the N-body covariance matrix is well reproduced by all approximate methods matching the initial conditions of Minerva (that is, all except the Lognormal case), with just slightly larger discrepancies from the \halogen estimate.

We obtain very similar results for Sample 2, with larger discrepancies
(roughly by a factor of two) for the \halogen and Lognormal
predictions.

%%%%%%%%%%%%%%%%%%%%%%%%%%%%%%%%%%%%%%%%%%%%%%%%%%%%%%%%
%%%%%%%%%%%%%%%%%%%%%%%%%%%%%%%%%%%%%%%%%%%%%%%%%%%%%%%%
\section{Comparison of the errors on cosmological parameters}
\label{sec:comperrors}

In addition to the direct comparison of bispectrum measurements and
their estimated covariance, we explore, as done in Papers I and II,
the implications for the determination of cosmological parameters of
the choice of an approximate method.

In this case we will consider a simpler likelihood analysis, compared
to those assumed for the 2-point correlation function and the power
spectrum in the companion papers. In the first place, the model for
the halo bispectrum, described in section~\ref{sec:model}, is a
tree-level approximation in PT and we will only consider its
dependence on the linear and quadratic bias parameters, along with two
shot-noise nuisance parameters. We also only consider the redshift-space
bispectrum monopole as the implementation and testing of
loop-corrections to the galaxy bispectrum in redshift space is well
beyond the scope of this work. 

In a first test, section~\ref{sec:constraints_var}, we will
  include in the likelihood only the estimate of the bispectrum
  variance, since 300 realisations are insufficient for any solid
  estimation of the covariance of more than 500 triangular
  configurations, as originally measured. In section
  \ref{sec:constraints_cov}, however, we consider a rebinning of these
  measurements that reduces the number of relevant configurations to
  less than a hundred and we will attempt a likelihood analysis
  involving the full bispectrum covariance. While the chosen
  wavenumber bin in this case is likely too large for a proper
  bispectrum analysis, it should allow, to some extent, comparison of
  different estimates of the bispectrum covariance matrix.  We will
  explore quantitatively the implications of limited number of
  realisations and the relative approximations in
  section~\ref{sec:tests}.

We will not consider any study of the cross-correlation between power
spectrum and bispectrum measurements, leaving that subject for future
work.

%%%%%%%%%%%%%%%%%%%%%%%%%%%%%%%%%%%%%%%%%%%%%%%%%%%%%%%
\subsection{Halo bispectrum model}
\label{sec:model}

We assume a tree-level model both for the matter bispectrum and the halo bispectrum. 

The real-space matter bispectrum $B_{m}$ is therefore given by \citep[see, \eg][]{BernardeauEtal2002}
\be
\label{eq:Bm}
 B_m(k_1,k_2,k_3) = 2 \,F_2(\kv_1,\kv_2) P_{m}^L(k_1) P_{m}^L(k_2) + \rm{2~perm.}\;\, 
 \ee
where $F_2$ is the quadratic PT kernel and $P_{m}^L(k)$ is the linear matter power spectrum. 

The halo bias model includes both local and nonlocal corrections
\citep{BaldaufEtal2012,ChanScoccimarroSheth2012,ShethChanScoccimarro2013} so that, at
second order, the halo density contrast takes the form
\be
\label{eq:brel}
\delta_h = b_1 \delta + \frac{b_2}{2} \delta^2 + \gamma_2\,\mathcal{G}_2, 
\ee
where $\mathcal{G}_2  $ is defined as
\be
\mathcal{G}_2 \equiv (\nabla_{ij} \Phi_v)^2 -  (\nabla^2\Phi_v)^2\,, 
\ee
with $\Phi_v$ being the velocity potential such that ${\bf v} = \nabla \Phi_v$. 

The full model for the real-space halo bispectrum therefore reads
\bea
\label{Bh}
B_{h}  & =  &  b_1^3 B_m(k_1, k_2, k_3) + \nn\\
& & +b_2\, b_1^2 \,\Sigma(k_1, k_2, k_3)   +\nn\\
& & + 2 \gamma_2 b_1^2 K(k_1, k_2, k_3)+\nn\\
& & +B_{SN}^{(1)}b_1^2 \left[P_m^L(k_1)+P_m^L(k_2)+P_m^L(k_3) \right]\nn\\
& & +B_{SN}^{(2)} \,,
\eea
where 
\be
\Sigma(k_1, k_2, k_3)  \equiv P_m^L(k_1)P_m^L(k_2) + 2~\rm cyc\,,
\ee
and 
\be
K(k_1, k_2, k_3)  \equiv
(\mu_{12}^2 -1) P_m^L(k_1)P_m^L(k_2) + 2\, \rm cyc\,,
\ee
$\mu_{12}$ being the cosine of the angle between $\kv_1$ and $\kv_2$. The last two
contributions account for any departure from the expected shot-noise
contribution under the Poisson assumption, see \eq{eq:Bshotnoise}. For
exactly Poisson shot-noise $B_{SN}^{(1)}=B_{SN}^{(2)}=0$ and we will
treat them here as free parameters with vanishing fiducial value.

Since we will consider the covariance for the {\em redshift-space} bispectrum, the corresponding model will be a slight modification accounting for the Kaiser effect on the power spectrum and  bispectrum monopoles. We will have then
\bea
\label{Bs}
B_{s}   & =  & a_0^B\left[ b_1^3 B_m(k_1, k_2, k_3) + \right.\nn\\
& & +b_2\, b_1^2 \,\Sigma(k_1, k_2, k_3)   +\nn\\
& & \left.+ 2 \gamma_2 b_1^2 K(k_1, k_2, k_3)\right]+\nn\\
& & +B_{SN}^{(1)}\,a_0^2\,b_1^2 \left[P_m^L(k_1)+P_m^L(k_2)+P_m^L(k_3) \right]\nn\\
& & +B_{SN}^{(2)} \,,
\eea
where, following \citet{ScoccimarroCouchmanFrieman1999, SefusattiEtal2006}, we model redshift-space effects on the bispectrum monopole simply in terms of the factor $a_0^B=1+2\beta/3+\beta^2/9$ with $\beta=f/b_1$, $f$ being the growth rate at $z=1$, while  $a_0=1+2\beta/3+\beta^2/5$ is the analogous factor for the power spectrum monopole, $P_s(k)=a_0\,P_h(k)$.  Such correction are not having any substantial effects on our results.

The model above will therefore depend on five parameters: the local
bias parameters $b_1$, $b_2$, the nonlocal bias parameter $\gamma_2$
and two shot noise parameters $B^{(1)}_{SN}$ and $B^{(2)}_{SN}$. We
will evaluate all matter correlators for our fiducial cosmology, along
with the growth rate $f$, and consider them as known in our analysis.

The fiducial values of $b_1$ for the two samples come from the
  comparison of the linear matter power spectrum and the halo power
  spectrum measured in the Minerva realisations.  The quadratic bias
  $b_2$ is in turn obtained from the linear one by means of the
  fitting formula in \cite{LazeyrasEtal2016} while for the non-local
  bias $\gamma_2$ we adopt the Lagrangian relation $\gamma_2 =
  -2/7(b_1-1)$ \citep{ChanScoccimarroSheth2012}.

We expect this model to accurately fit simulations over a quite small
range of scales, typically for $k<0.07 \kMpc$ \citep[see,
  \eg][]{SefusattiCrocceDesjacques2012, BaldaufEtal2015A,
  SaitoEtal2014}. However, we assume it to represent a full model for
the halo bispectrum down to $0.2\kMpc$ since we are merely interested
in assessing the relative effect of different estimate of the variance
on parameter determination. The value of $k_{max}=0.2\kMpc$ is,
nevertheless, a reasonable estimate of the reach of analytical models,
once loop corrections are properly included
\citep{BaldaufEtal2015A}. We leave a more extensive investigation,
including a joint power spectrum-bispectrum likelihood to future work.

%%%%%%%%%%%%%%%%%%%%%%%%%%%%%%%%%%%%%%%%%%%%%%%%%%
\subsection{Likelihood}
\label{sec:likelihood}

We assume a Gaussian likelihood for the  bispectrum given by
\be
\ln \mathcal{L}_{B}=-\frac12 \sum_{ij}\delta B_i\,\left[C\right]^{-1}_{ij}\,\delta B_j\,,
\label{eqlnL}
\ee
where $\d B\equiv B_{data}-B_{model}$ while $C_{ij}$ is the bispectrum covariance matrix with indices $i$ and $j$ denoting individual triangular configurations $t_i$. The sum runs over all triangular configurations, $i=1,\dots,N_t$, $N_t$ being their total number corresponding to a chosen value for the smallest scale included in the analysis and determined by the parameter $k_{max}$. This is given by
\be
\label{eq:Nt}
N_t=\sum_{k_1=\Delta k}^{k_{max}}\sum_{k_2=\Delta k}^{k_1}\sum_{k_3=\max(\Delta k,k_1-k_2)}^{k_2} 1\,
\ee
where the sums ensures that $k_1\ge k_2\ge k_3$ and that all triangle
bins include closed fundamental triangles. Such quantity can be computed analytically, albeit in terms of ceiling and floor functions, as shown in \citet{ChanBlot2017}. As mentioned above, assuming the $k$-bin $\Delta k =
  3 k_f$ adopted for the original measurements and $k_{max}=0.2\kMpc$ we obtain $N_t=508$. In section~\ref{sec:constraints_cov} we will consider a re-binning of all triangular configurations assuming $\Delta k = 6 k_f$ and the same value for $k_{max}$ leading to $N_t =82$.

Similarly to the analyses in Paper I and II, since we are not
interested in evaluating the accuracy of the model we assume, but only
to quantify the {\em relative} effect of replacing the variance
estimated from the N-body realisations with those obtained with the
approximate methods, we assume as ``data'' the ``model'' bispectrum
evaluated at some fiducial values for the parameters, that is
$B_{data}=B_{model}(p_\alpha^{*})$. While this lead to a vanishing
$\chi^2$ for the best fit/fiducial values, it still allows to estimate
how the error on the parameters depends on the bispectrum covariance
estimation.

Our choice for the parameters allows to obtain an analytical dependence of the likelihood function on them, that does not require a MonteCarlo evaluation. In fact, we can rewrite the model in \eq{Bh} as
\be
B_{model}   =  \sum_{\alpha=1}^5 p_\alpha\, \mathcal{B}_\alpha\,,
\ee
where  $\left\{p_{\alpha}\right\}=\left\{ a_0^B\,b_1^3, a_0^B\,b_1^2\,b_2 \,, a_0^B\,b_1^2\,\gamma_2\,, a_0^2\,b_1^2 \,B_{SN}^{(1)}\,, B_{SN}^{(2)} \right\}$ and $\left\{\mathcal{B}_\alpha\right\}  = \left\{ B_m , \Sigma, 2\,K, P_m(k_1)+P_m(k_2)+P_m(k_3), 1\right\}$.
Given our specific definition of $B_{data}$ we can also write
\be
-\d B=B_{model}-B_{data}   =  \sum_{\alpha=1}^5 (p_\alpha-p_\alpha^*)\, \mathcal{B}_\alpha\,,
\ee
and therefore it is easy to see that we can rewrite the likelihood as
\be
\ln \mathcal{L}_{B}  =  
-\frac12  \sum_{\alpha,\beta=1}^{5}\, (p_\alpha-p_\alpha^*)\,(p_\beta-p_\beta^*)\,\mathcal{D}_{\alpha\beta}\,,
\ee
where
\be\label{eq:D}
\mathcal{D}_{\alpha\beta}\equiv \sum_{i,j=1}^{N_t}\,\mathcal{B}_{\alpha}(t_i)\,\left[C^B\right]^{-1}_{ij}\,\mathcal{B}_{\beta}(t_j)\,.
\ee
In this way the likelihood $\mathcal{L}_{B}$ is explicitly written as
an exact, multivariate Gaussian distribution in the parameters
$p_\alpha$. Clearly, once the quantities $\mathcal{D}_{\alpha\beta}$
are computed, we can evaluate any marginalisation analytically. We
could, in principle consider a transformation between these parameters
and the set given by $\left\{ b_1,b_2, \gamma_2, B_{SN}^{(1)},
B_{SN}^{(2)} \right\}$ but this would require an approximation for the
likelihood around its maximum and, furthermore, it would not add any
information to our goal since any relative variation on the error on
the parameter cube $b_1^3$, for instance, is of the same order as the
relative variation on the error on $b_1$. We refer the reader to
  \citet{2017MNRAS.471.1581B} for a recent analysis in terms of
  cosmological parameters of the matter bispectrum and several other
  related observables.
  
%%%%%%%%%%%%%%%%%%%%%%%%%%%%%%%%%%%%%%%%%%%%%%%%%%%%%%%%%%%%%%%%%%
\subsection{Constraints comparison: variance}
\label{sec:constraints_var}

As a first test, we consider the comparison of the errors on the
  parameters obtained from the bispectrum variance estimated for the
  triangular configurations defined by the $k$-bin $\Delta k = 3
  k_f$. As already mentioned, even restricting our analysis to
  $k_{max}=0.2\kMpc$, we end up with $N_t=508$ triangles, a number
  larger than the total number of realisations at our disposal,
  precluding a robust estimate of the covariance matrix. In this
  section, therefore, we approximate \be
  \mathcal{D}_{\alpha\beta}\simeq
  \sum_{i=1}^{N_t}\,\frac{\mathcal{B}_{\alpha}(t_i)\,\mathcal{B}_{\beta}(t_j)}{\Delta
    B^2(t_i)}\,, \ee $\Delta B^2(t_i)$ representing the variance for
  the triangular configuration $t_i$.

\begin{figure*}
\begin{center}
\includegraphics[width=0.95\textwidth]{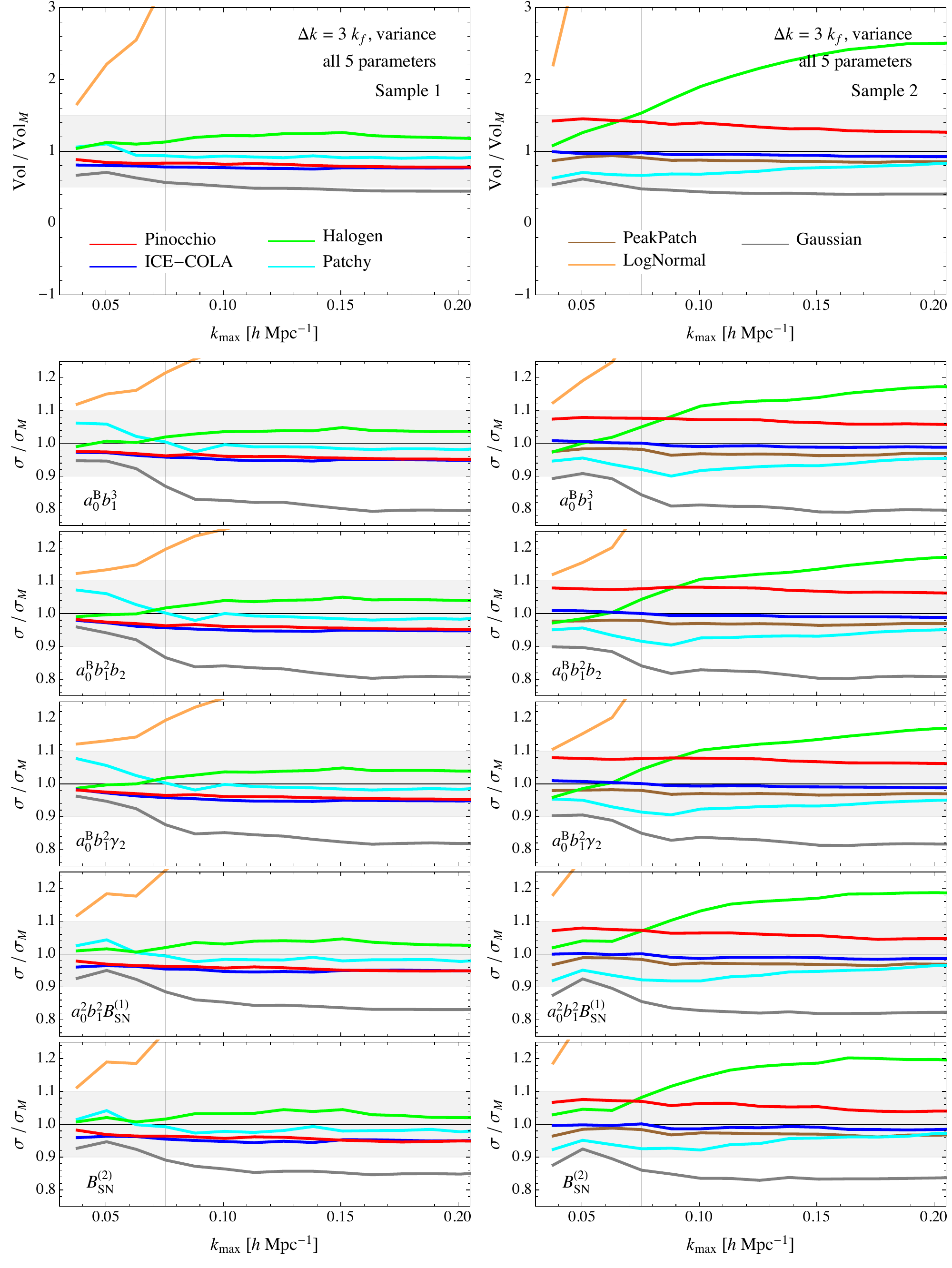}
\caption{\label{fig:margErrRSD1D2} Marginalized errors for the model
  parameters obtained in terms of the redshift space bispectrum
  variance estimated with approximate methods compared with the errors
  obtained from the N-body estimate of the variance. First and second
  column correspond respectively to Sample 1 and Sample 2. See text
  for an explanation. }
\end{center}
\end{figure*}

Figure~\ref{fig:margErrRSD1D2} shows the ratio between the
marginalised error on each parameter $p_\alpha$ obtained from the
variance estimated with a given approximate method and the same
marginalised error on the same parameter obtained from the variance
estimated from the Minerva N-body set. Such ratio is shown as a
function of the maximum wavenumber $k_{max}$ assumed for the
likelihood evaluation that defines as well the total number of
configurations $N_t$ according to \eq{eq:Nt}. The left column
corresponds to Sample 1 while the right column to Sample 2. The grey
shaded area represents a 10$\%$ discrepancy between error estimates.

In addition to the errors on individual parameters we consider, as in
the companion papers, the volume of the 5-dimensional ellipsoid
corresponding to the combined errors on all parameters defined as
\be
\label{eq:vol}
{\rm Vol} = \sqrt{\det {\mathcal D}_{\alpha\beta}^{-1}}\,,
\ee
since ${\mathcal D}_{\alpha\beta}^{-1}$ represents the parameters
covariance matrix. The ratio of this quantity estimated from the
approximate methods and from the N-body runs is shown in the two top
panels of figure~\ref{fig:margErrRSD1D2} for the two samples. In this
case, the shaded area corresponds to a discrepancy of 50\%, reflecting
the target 10\% for individual parameters.

These results reflect those shown in the comparison of the
variance. Unsurprisingly the methods that overestimate the variance
lead to an overestimate of the error on each parameter, in a similar
fashion across all parameters. As already shown in the previous
figures, the predictive methods, along with \patchy, appear to be more
accurate, with \cola, in particular, the one providing more consistent
results for both samples. All such methods show discrepancies of less
than 10\% w.r.t. the N-body case. The behaviour of \halogen is also
quite good in the low-mass sample but the difference with N-body
becomes larger than 10$\%$ in the second sample once small scales are
included. LogNormal shows the largest difference, with reasonable
results only for the very large scales.  The Gaussian theoretical
  prediction provides a reasonable estimate at the largest scales
  while it underestimate the variance at small scales, particularly in
  the case of the parameters more directly related to bias, probably
  due to a missing non-Gaussian component.

\begin{figure*}
\begin{center}
\includegraphics[width=0.95\textwidth]{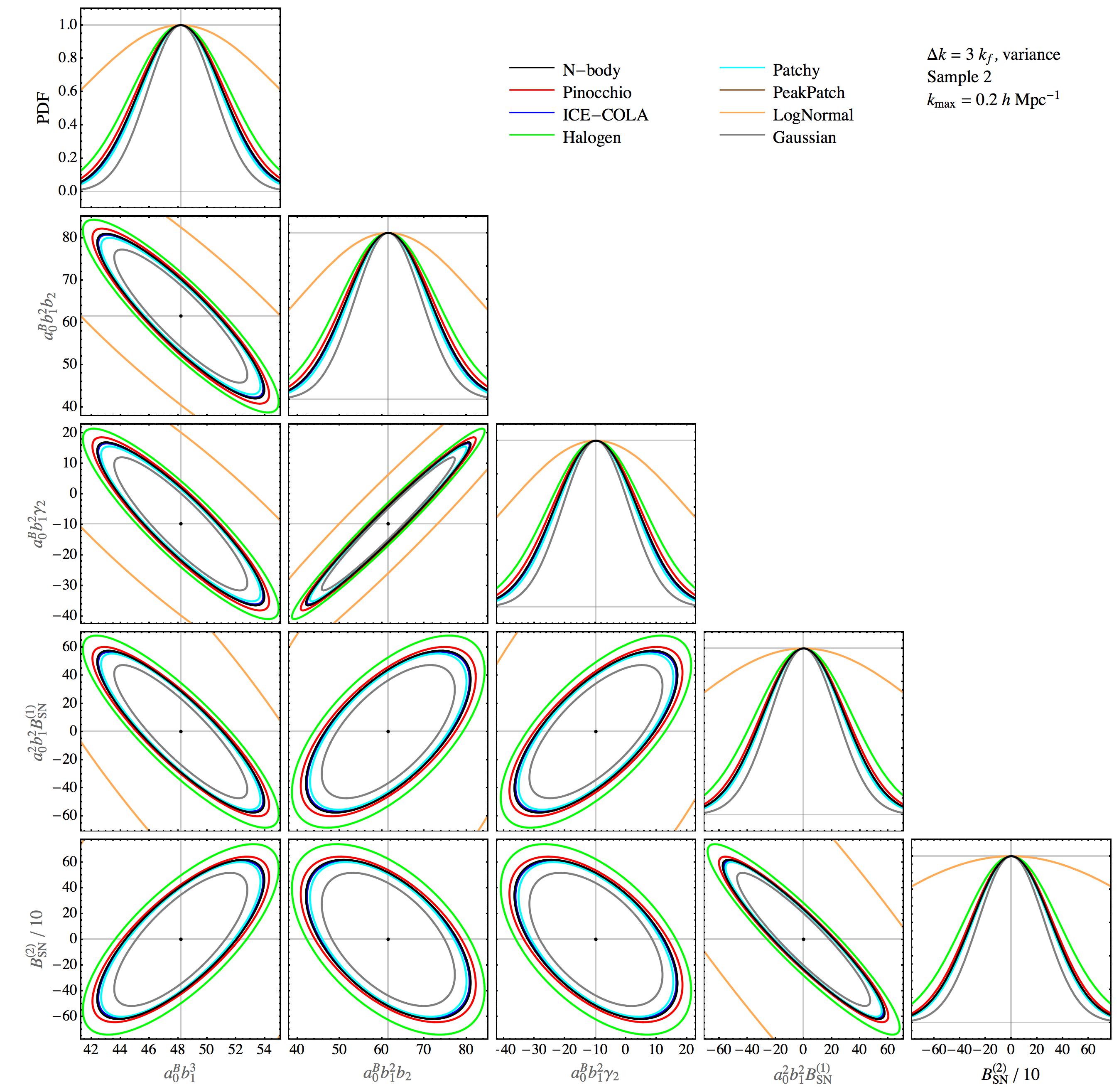}
\caption{\label{fig:contRSD2} 2-$\sigma$ contour plots for the
  parameters combinations $p_\alpha$ (see text) from the bispectrum
  monopole in redshift space for Sample 2. The constraints assume
  $\kmax$=0.2$\kMpc$. Notice that the N-body (black) results are
  plotted on top so that a few curves, corresponding to methods very
  closely reproducing the N-body, ones are not easily visible. }
\end{center}
\end{figure*}

Finally, figure~\ref{fig:contRSD2} as an example, shows the 2-$\sigma$
contour plots for the parameters combinations $p_\alpha$ in redshift
space. Similar results are obtained for Sample 1.  One can notice, in
particular, that no method provides a variance estimate that affects
the degeneracies between parameters in any specific way. Such effect
might be more relevant when the full covariance is taken into
account. We will comment on this in the next section.

%%%%%%%%%%%%%%%%%%%%%%%%%%%%%%%%%%%%%%%%%%%%%%%%%%%%%%%%%%%%%%%%%
\subsection{Constraints comparison: covariance}
\label{sec:constraints_cov}

In this section we consider a comparison of the recovered
  parameters errors that accounts for the full covariance among
  triangular configurations. Clearly we need to reduce significantly
  the total number of triangles in order to recover reliable estimates
  of the covariance matrix even with our small set of independent
  realisations. We do so by rebinning our measurements in triangular
  configurations with sides defined by $k$-bins of size $\Delta k = 6
  k_f=0.025\kMpc$. This is quite a large value leading to triangular
  bins, each accounting for a large set of fundamental triangles of
  quite different shape. For this reason it is probably not a good
  choice for a proper bispectrum analysis that aims at taking
  advantage of the different shape-dependence of the various
  contribution to the galaxy bispectrum. However, it can still provide
  a reasonable estimate of the covariance matrix, at least in the
  context of our comparison with N-body simulations. 

As already mentioned, this binning choice leads to a total number
  of triangles of at most $N_t=86$ for $k_{max}=0.2\kMpc$. We will then
  assume that their covariance matrix can be estimated reasonably well
  from the relatively small number realisations available for each method and we can
  therefore compute the likelihood function in terms of \eq{eq:D}.  We
  do this for all possible values of $k_{max}$ in steps of $\Delta k$. Notice that, despite the reduced number of triangular configurations, we correct the parameters covariance matrix by the factor shown in eq. (18) of
  \cite{PercivalEtal2014} to take into account the uncertainty in the estimated inverse covariance.
  
\begin{figure*}
\begin{center}
\includegraphics[width=0.95\textwidth]{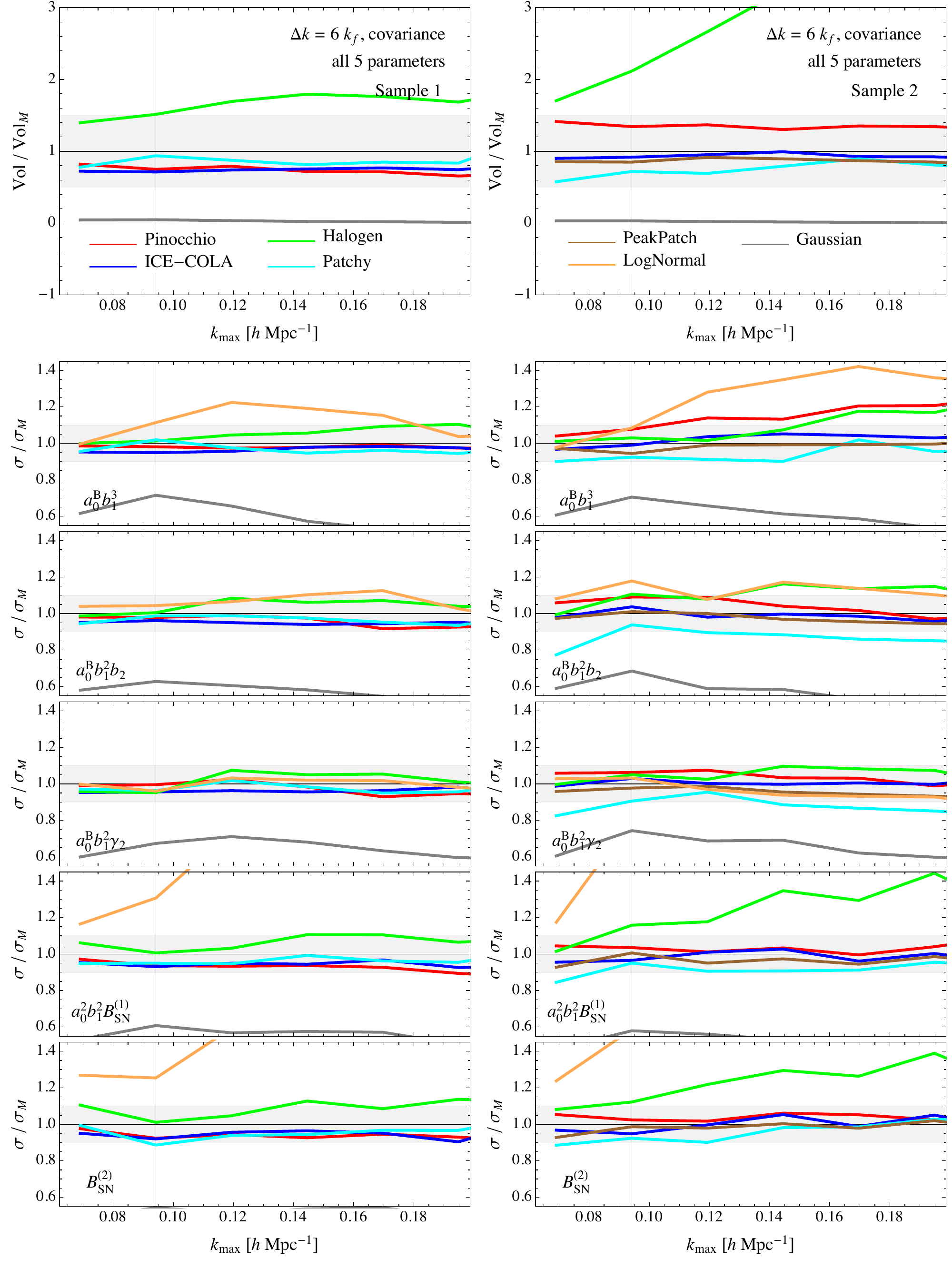}
\caption{\label{fig:margErrRSD1D2cov} Same as
  figure~\ref{fig:margErrRSD1D2} but assuming a $k$-bin of size
  $\Delta k=6 k_f$ and the full covariance matrix for all triangular
  configurations.}
\end{center}
\end{figure*}

  The comparison of the individual
  marginalised errors is shown in
  figure~\ref{fig:margErrRSD1D2cov}. They are not too different from
  the previous one to the extent that most methods, and the predictive
  ones in the first place, do lead to errors within 10\% of those
  obtained from the N-body-based covariance on individual
  parameters. One can notice, however, a somehow larger discrepancy in
  the case of \halogen and a much larger one for the LogNormal
  estimate which is out of the shown interval in the case of the
  5-parameter volume comparison. On the other hand, the Gaussian
  theoretical prediction is responsible for an even more significant
  underestimate of the errors with respect to the previous case, as
  one can expect, at least in part, since it constitutes a diagonal
  approximation for the full covariance matrix.

\begin{figure*}
\begin{center}
\includegraphics[width=0.95\textwidth]{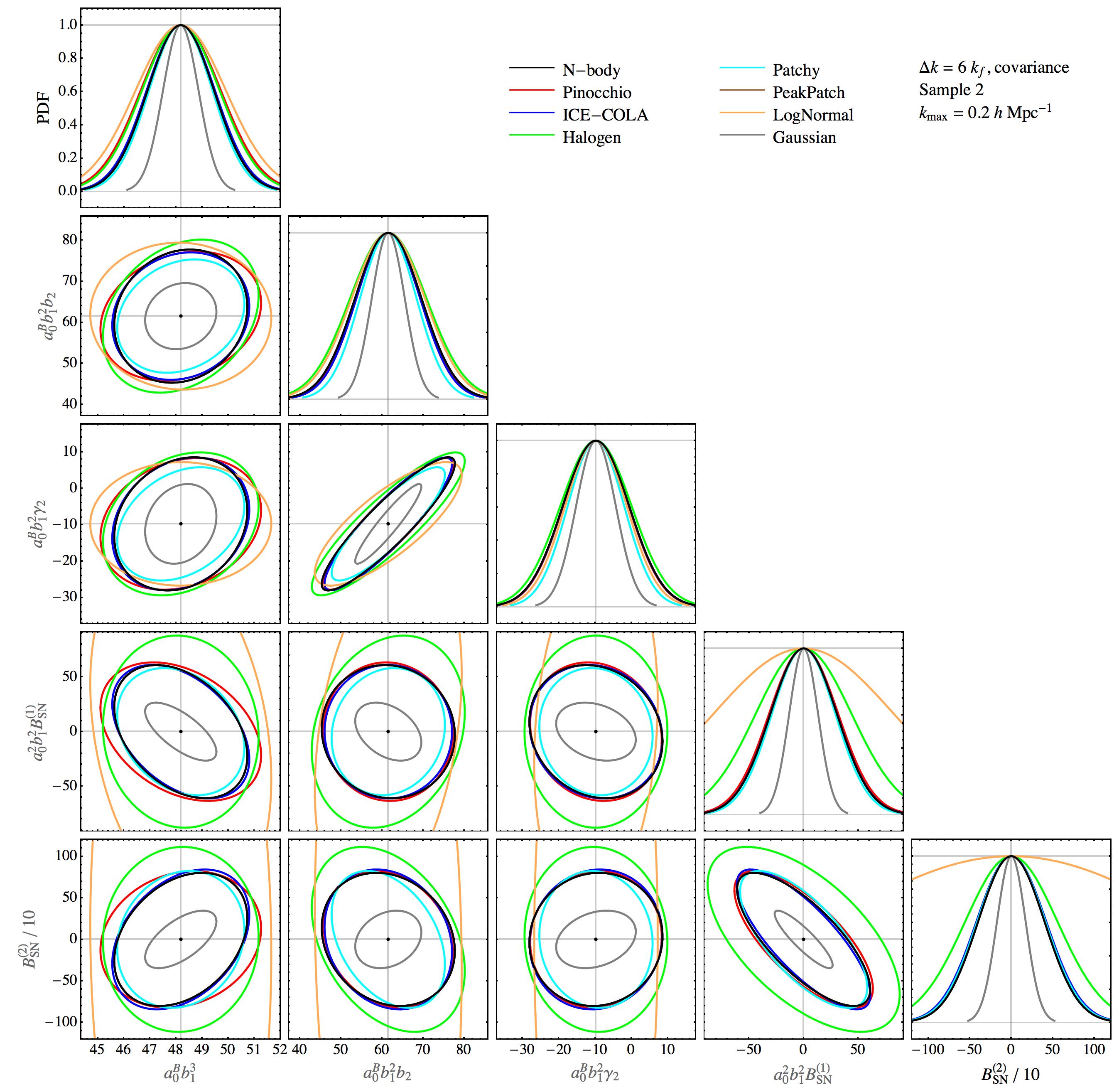}
\caption{\label{fig:contRSD2cov} Same as figure \ref{fig:contRSD2} but
  assuming a $k$-bin of size $\Delta k=6 k_f$ and the full bispectrum
  covariance matrix}
\end{center}
\end{figure*}

Figure~\ref{fig:contRSD2cov} shows the marginalised 2-$\sigma$
  contours for pairs of parameters in the case of Sample 2 and
  $k_{max}=0.2\kMpc$. In addition to the considerations just made, one
  can observe how, in the context of the covariance comparison,
  different methods lead to slightly different degeneracies among the
  parameters, something not evident in the previous case of the
  variance comparison. \halogen, LogNormal and the Gaussian (diagonal)
  prediction, in fact, stand out not only for the larger/smaller
  errors bars recovered but also for the degeneracies they provide for
  some couples of parameters.

%%%%%%%%%%%%%%%%%%%%%%%%%%%%%%%%%%%%%%%%%%%%%%%%%%%%%%%%%%%%%%%%%
%%%%%%%%%%%%%%%%%%%%%%%%%%%%%%%%%%%%%%%%%%%%%%%%%%%%%%%%%%%%%%%%%
\section{Tests with a large set of realisations}
\label{sec:tests}

The number of 300 realisations, despite being quite a large number for
many applications, is still rather small when it comes to estimate the
covariance of hundreds or thousands of bispectrum
configurations. For this reason we limited our likelihood
  comparisons to the bispectrum variance alone or, as an alternative,
  we used very large bins of wavenumber to reduce the number of
  triangles.

In this section we test the robustness of some of our conclusions
taking advantage of a much larger sets of 10,000 \pin catalogs
characterised by the same configuration and cosmology as the 300 so
far considered. In particular this will allow us to investigate
  the relevance of the off-diagonal elements of the bispectrum
  covariance matrix for the two different binnings.

\begin{figure*}
\begin{center}
\includegraphics[width=0.95 \textwidth]{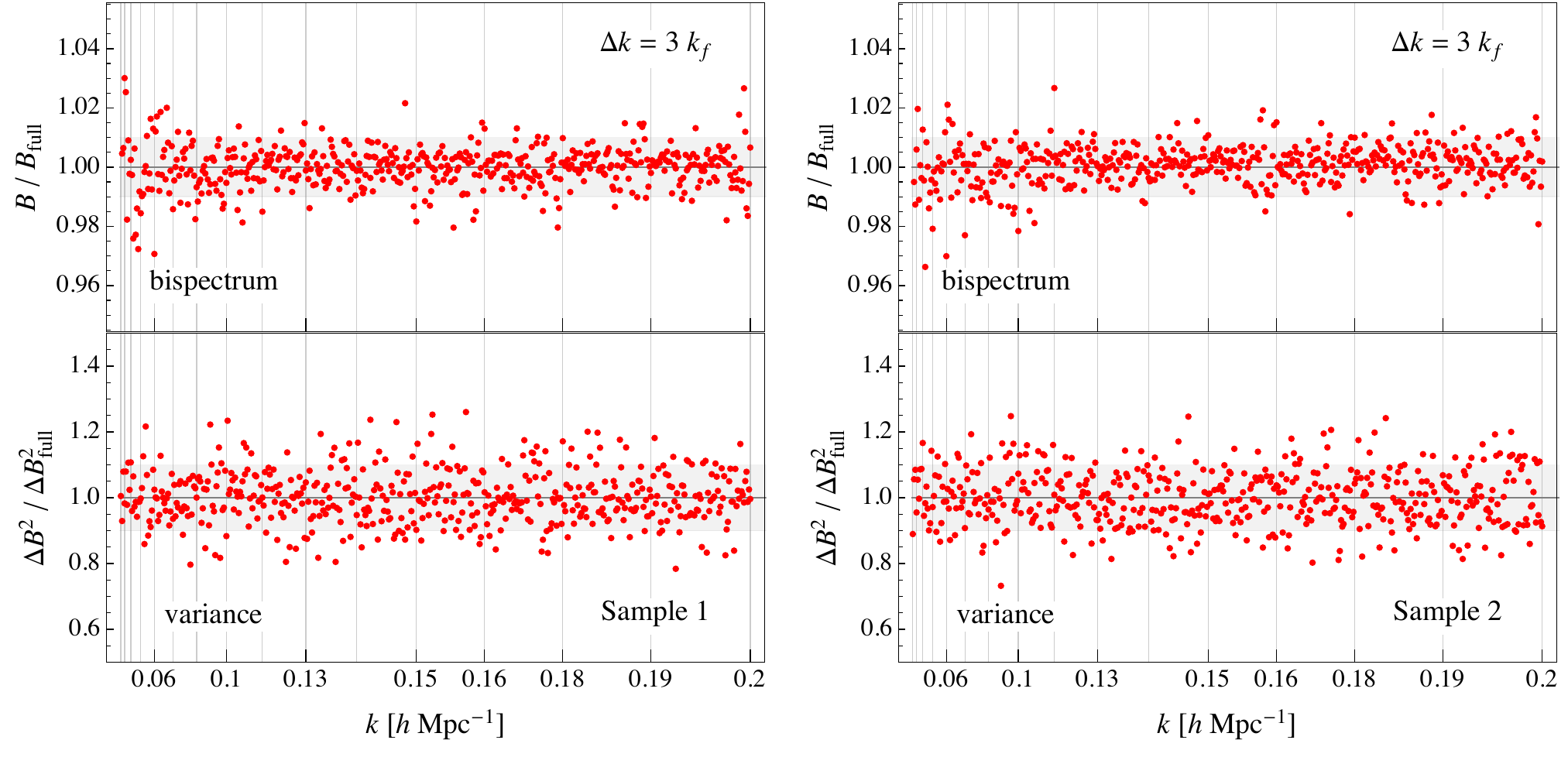}
\includegraphics[width=0.95 \textwidth]{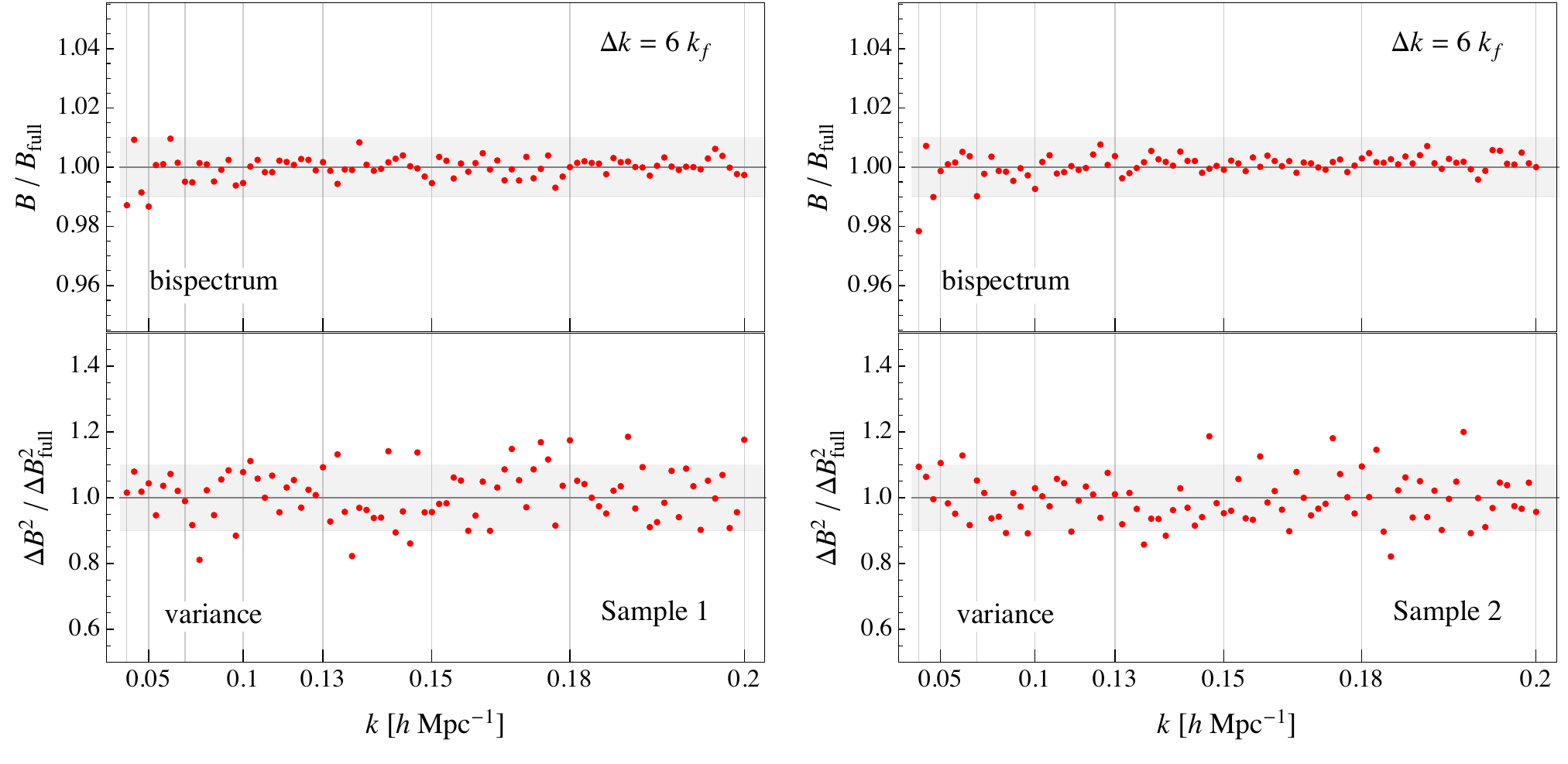}
\caption{\label{fig:bsRealAll1e5D1D2} Ratio between the
    bispectrum and its variance as measured in 300 realisations of
    \pin to the same quantities estimated from 10,000 realisations in
    real space. Top panels assumes $\Delta k = 3 k_f$, bottom panels
    $\Delta k = 6 k_f$. Sample 1 and Sample 2 are shown respectively
    in the left and right column.} 
\end{center}
\end{figure*}

In figure \ref{fig:bsRealAll1e5D1D2} we show the ratios of the
  real-space bispectrum and its variance obtained from 300
  realisations and the same quantities obtained from the 10,000 runs
  for both samples and for both binning choices. In the small-bin case
  the scatter on the bispectrum due to the limited number of runs is
  of the order of a few percent, while for the variance is of the
  order of 10\%, with no particular dependence on shape. For the large
  bin the scatter is reduced below 1\% for the bispectrum and about at
  that level for its variance. Such differences are smaller than the
  discrepancies discussed among the results from different methods in
  the previous sections. 

We look now at the effect of poor sampling on the off-diagonal
  elements of the covariance matrix in terms of the cross-correlation
  coefficients defined as \be
  r_{ij,full}\equiv\frac{C_{ij}}{\sqrt{C_{ij,full}C_{jj,full}}} \,,
  \ee where $C_{ij,full}$ represents the covariance between triangles
  $t_i$ and $t_j$ estimated from the 10,000 runs, while the $C_{ij}$
  in the numerator represents the covariance from only 300
  realisations. Comparing this quantity with the cross-correlation
  coefficients estimated fully from the 10,000 runs allows to identify
  discrepancies directly as differences between the two covariance
  matrices.
  
\begin{figure*}
\begin{center}
\includegraphics[width=0.95\textwidth]{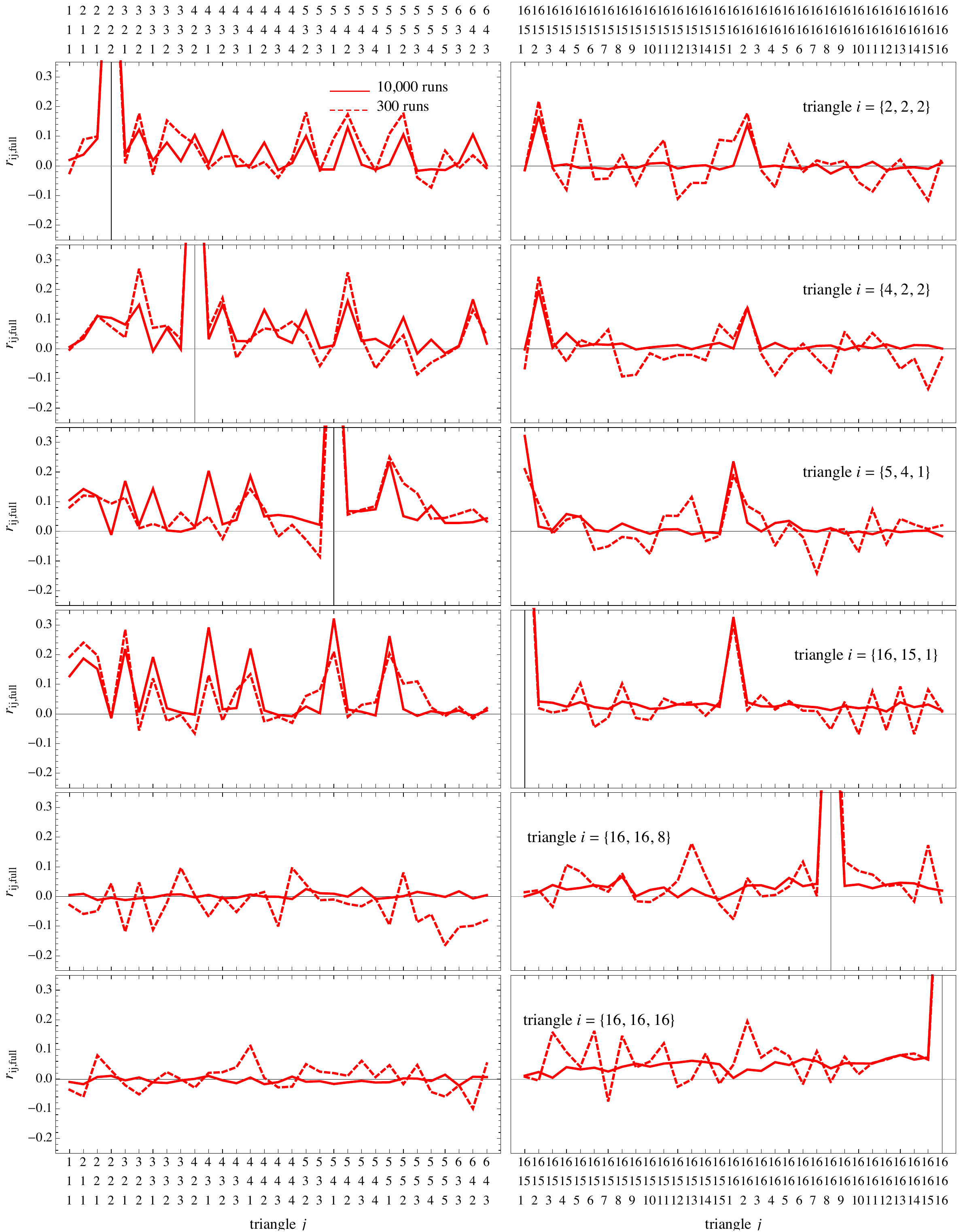}
\caption{\label{fig:bsRealccM1e5rowsD1} Cut through the
  cross-correlation coefficient in real space for all the triangle
  configurations coefficients estimated from 300 realizations (dashed
  line) and 10,000 realizations (continuous line) for the first
  sample. On the x-axis there are the triplets for each triangle in
  fundamental frequency unit.  The cross-correlation coefficient is
  normalized to the Minerva variance.}
\end{center}
\end{figure*}

  Figure~\ref{fig:bsRealccM1e5rowsD1} shows a selection of
  elements from the $r_{ij,full}$ matrix for the measurements assuming
  the bin $\Delta k = 3 k_f$ and Sample 2.  The two subsets of
configurations on the abscissa correspond to triangles at the largest
and smallest scales considered (respectively left and right column)
under the assumption of $k_{max}=0.2\kMpc=16 \Delta k$. It is
interesting to notice how the noise characterising the first estimates
is of the order of the true off-diagonal correlations from the less
noisy estimate, present, as expected, between configurations sharing
one or two wavenumbers, \eg~$t_i=\left\{2,2,2\right\}$ and
$t_j=\left\{16,15,2\right\}$.

\begin{figure*}
\begin{center}
\includegraphics[width=0.95\textwidth]{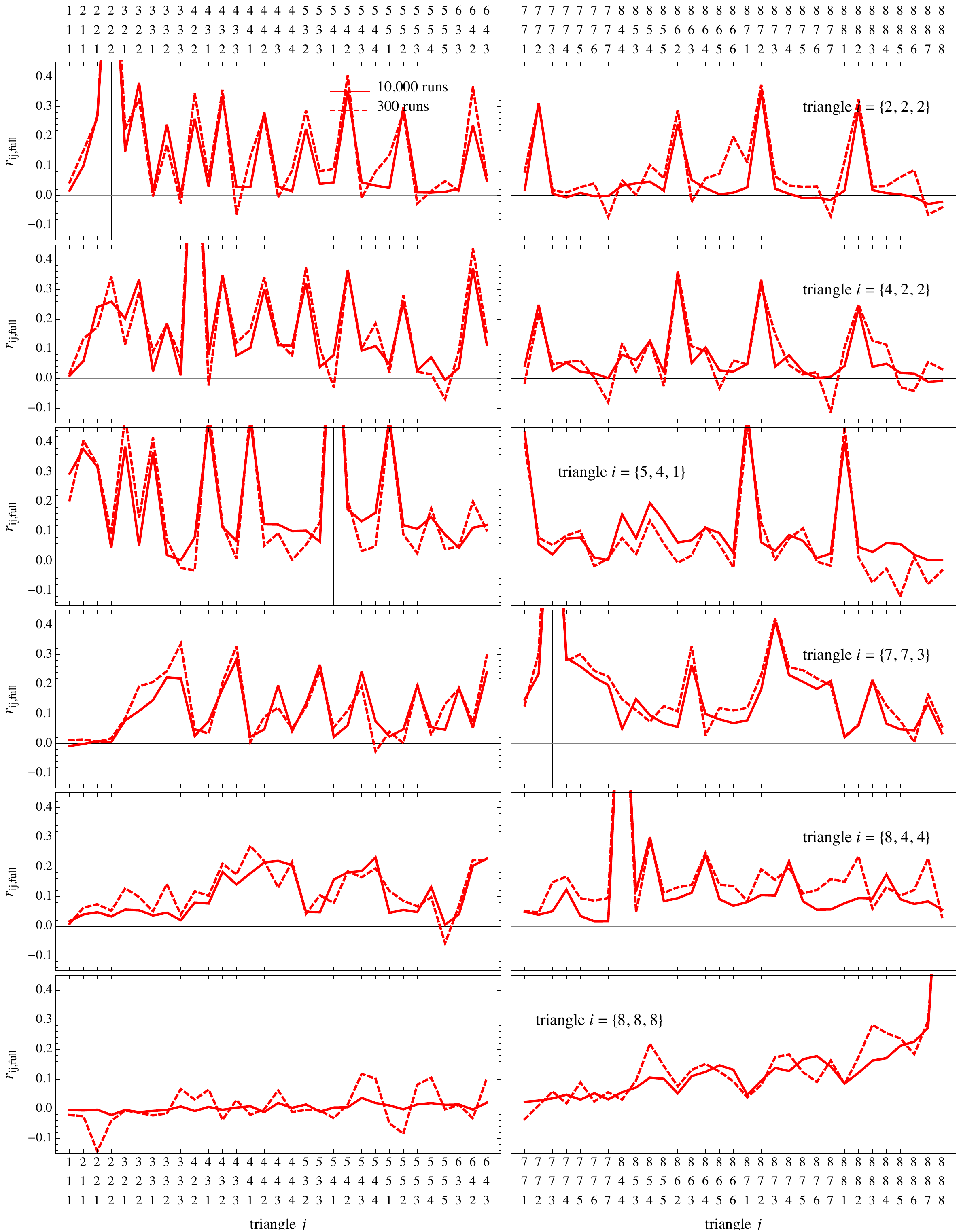}
\caption{\label{fig:bsRealccM1e5rowsD16kF} Same as figure
  \ref{fig:bsRealccM1e5rowsD1} but with $\Delta k = 6 k_f$.}
\end{center}
\end{figure*}

Figure~\ref{fig:bsRealccM1e5rowsD16kF} shows the same
  cross-correlation coefficients but for the larger binning, $\Delta k
  = 6 k_f$. Also in this the small-scale set of triangles correspond
  to the configurations close to the limit of $k_{max}=0.2\kMpc=8
  \Delta k$. The main difference with the previous case is the larger
  level of the correlations in the off-diagonal elements, due to the
  increased number of shared sides in the fundamental triangles
  falling in the larger triangular bins. The difference between the
  estimates from the 300 and 10,000 realisations sets is, however,
  smaller; in this case the off-diagonal structure of the covariance
  matrix is broadly reproduced even with 300 realisations, so this can
  be taken as a confirmation of the validity of the tests presented
  above.

\begin{figure*}
\begin{center}
\includegraphics[width=0.95\textwidth]{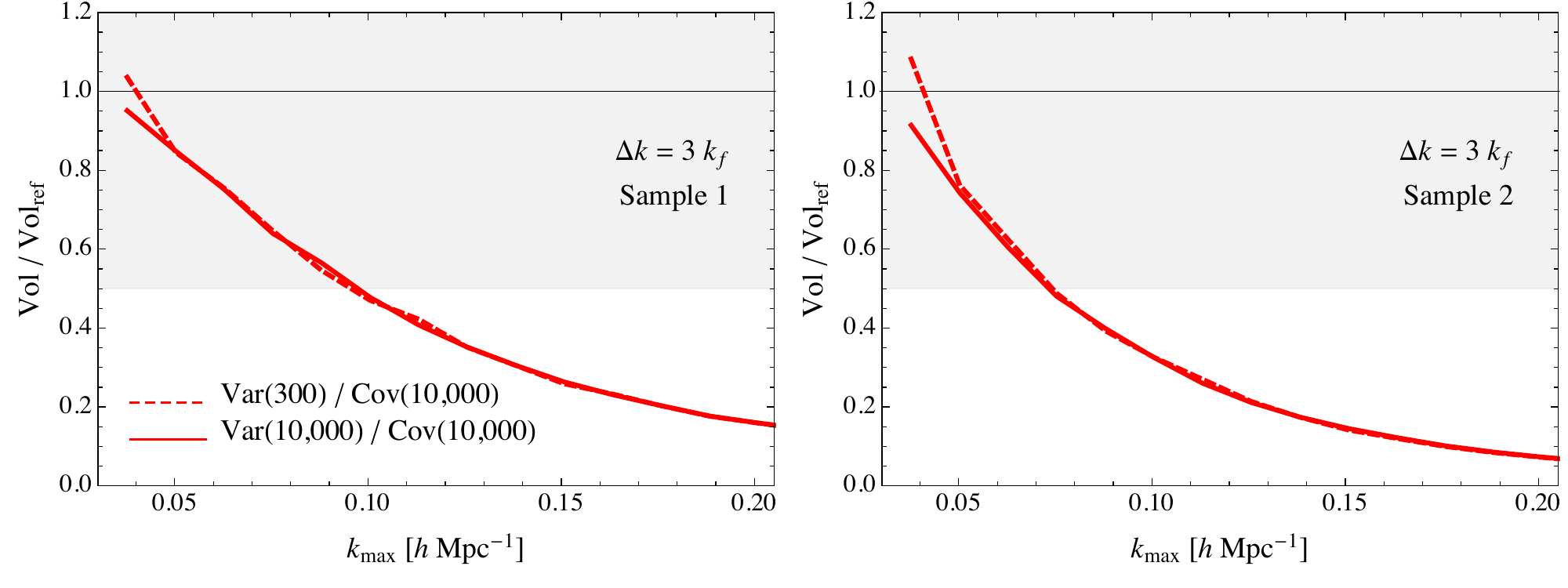}
\includegraphics[width=0.95\textwidth]{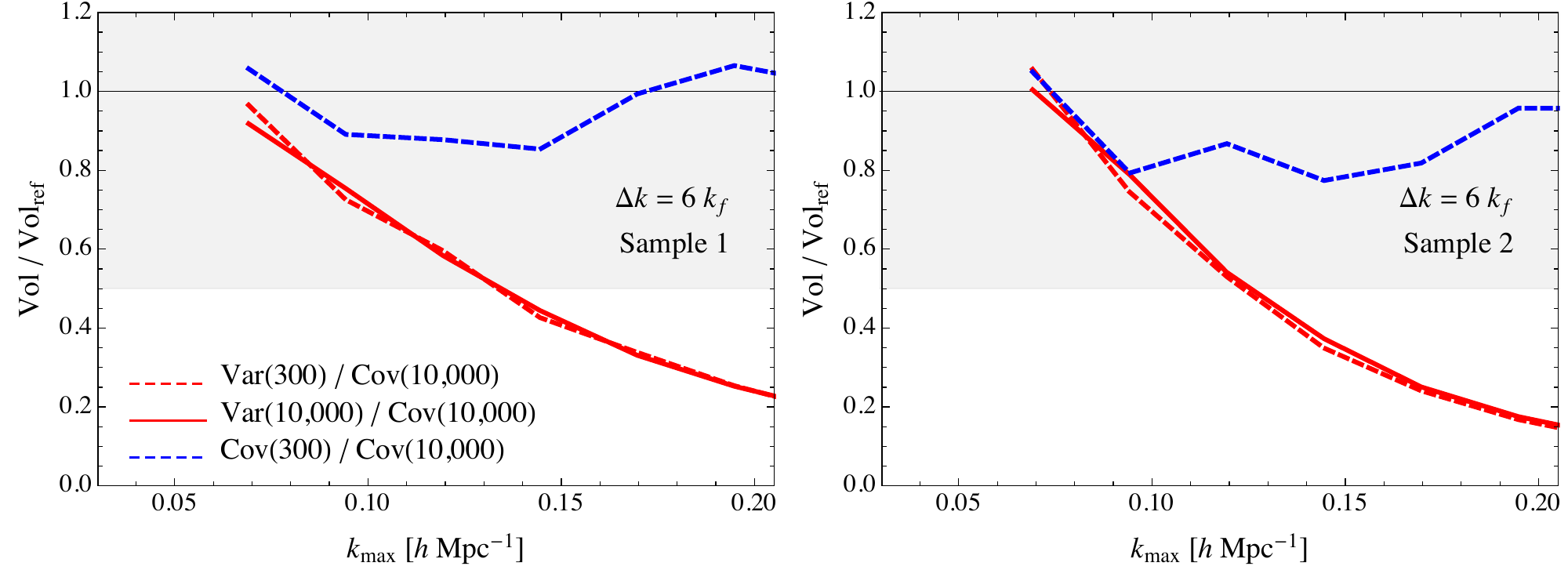}
\caption{\label{fig:margErrReal1e5D1D2} Errors volume, as defined
    in \eq{eq:vol} for the bias parameters; in the top panel the case
    $\Delta k = 3 k_f$ with the bispectrum variance from 10,000
    realizations (continuous line) and the variance from 300
    realizations (dashed line) compared with variance or the full
    covariance from 10,000; in the bottom panel the same but in the
    case $\Delta k = 3 k_f$ with the additional dashed line accounting
    for the comparison between the full covariance matrix from 300
    realisations and the full covariance matrix from 10,000
    realisations. In the first column the results are shown for the
    first sample, in the second column for the second sample.}
\end{center}
\end{figure*}
 
Finally, the top panels of figure~\ref{fig:margErrReal1e5D1D2} show the
    comparison between the volume error as defined in \eq{eq:vol} for
    the 5 parameters $p_\alpha$ obtained from the bispectrum variance
    estimated with 300 realisations, ``Var(300)'' (the case adopted
    for the results in section~\ref{sec:constraints_var}) and with the
    full set of 10,000 realisations, ``Var(10,000)'' against the same
    quantity derived in terms of the covariance from all 10,000 runs,
    ``Cov(10,000)''. We notice, in the first place, that no difference
    is noticeable in the results obtained from the variance estimated
    from the small or full sets. The difference between these and the
    case of the full covariance matrix is instead quite significant at
    almost all scales, except the very largest.  In particular, for
    $k_{max}=0.2\kMpc$, the analysis based on the full covariance
    provides an error volume almost an order of magnitude larger
    w.r.t. the variance one, although, at the level of the
    marginalised errors on individual parameters (not shown) the
    difference is of the order of 10\%, i.d. comparable to the
    difference among the different methods.

Similar results are shown, in the bottom panels of
  figure~\ref{fig:margErrReal1e5D1D2}, for the large binning $\Delta
  =6k_f$.  In the case we can compute as well the covariance from the
  restricted set of 300 realisations, ``Cov(300)''. The volume error
  in this case is still significantly smaller from the reference case
  by a 50\% at $k_{max}=0.2\kMpc$, but less than in the variance-based
  cases. As already done in section \label{sec:constraints_cov}, for this last comparison, we
  correct the parameters covariance in the ``Cov(300)'' case by the factor shown in eq. (18) of
  \cite{PercivalEtal2014} to take into account the small number of
  realisations used to estimate the covariance matrix when compared
  with the error measured from 10,000.

%%%%%%%%%%%%%%%%%%%%%%%%%%%%%%%%%%%%%%%%%%%%%%%%%%%%%%%%%%%%%%%%%%
%%%%%%%%%%%%%%%%%%%%%%%%%%%%%%%%%%%%%%%%%%%%%%%%%%%%%%%%%%%%%%%%%%
\section{Conclusions}
\label{sec:conclusions}

In this paper, and in its companions Papers I and II, we have studied
the problem of covariance matrix estimation for large-scale structure
observables using dark matter halo catalogs produced with approximate
methods. This last paper, in particular, focuses on the halo
bispectrum and its covariance matrix, with the twofold aim of
assessing the correct reproduction of the non-Gaussian properties of
the halo distribution as well as considering the halo/galaxy
bispectrum as a direct observable in its own right.
  
The measurements are performed on sets of 300 (1000 for LogNormal)
catalogs obtained from several different methods: \cola,
\peak, \pin, \halogen, \patchy, LogNormal and they are
compared with the reference Minerva suite of 300 N-body
simulations. All approximate catalogs, apart from LogNormal, assume
the same initial conditions of the full N-body simulations, thereby
reducing differences due to cosmic variance. Out of each halo
catalog we select two samples characterised by a different minimal
mass in order to gain a better perspective on our results as a
function of mass and shot-noise levels.
  
The approximate methods can be generically subdivided into predictive
methods (\cola, \pin, \peak), requiring a single redefinition of the
halo mass to recover the expected halo number density, and methods
(\halogen, \patchy), requiring as well a calibration of the bias
function. It should be noted that, in the case of \halogen, such bias
calibration is limited to the 2-Point Correlation Function and to
configuration space, with only one parameter (per mass-bin)
controlling the clustering amplitude. In addition, a third type is
represented by the Lognormal method, relying on a non-linear
transformation of the matter density field, in turn calibrated on the
halo mass function and halo bias. In all our analysis (with the
exception of Appendix~\ref{app:masscuts}) we have changed the limiting
mass for each sample in order to ensure the same abundance for all
catalogs, including those obtained with more predictive methods.

We have shown that:
\vspace{-2.8mm}
\begin{enumerate}
\item
  the real space bispectrum is reproduced by \cola, \pin, \patchy and
  \peak within 20$\%$ for the most of the triangle configurations
  while \halogen and, particularly, Lognormal present larger disagreements, often beyond 50\%; 
  \item
these discrepancies are reflected on the results for the bispectrum
variance, where, however, their systematic nature is less evident
since there is no clear dependence on the triangle shape, probably due
to the fact that for most triangles, the variance is dominated by the
shot-noise component; the Gaussian prediction for the variance is
generically underestimating the N-body result, particularly for
squeezed triangles;
\item
similar conclusions can be made for the redshift-space bispectrum
monopole, where, however, \patchy and \halogen (the latter at least for
the small mass sample) show a better agreement with the N-body
simulations;
\item
the inspection of the cross correlation coefficients illustrates how,
due to the matching initial conditions, almost all methods (except
Lognormal by construction) reproduce the noise present in the N-body
estimation, which is dominating the off-diagonal elements of the
covariance matrix estimated from only 300 realisations.
\end{enumerate}
  
Our analysis was not limited to how accurately the bispectrum and
  its covariance are recovered, but include a comparison of the errors
  on cosmological parameters, in this case linear and non-linear bias
  parameters, derived from each approximate estimate of the both the
  variance and covariance of the halo bispectrum in redshift
  space. Since the relatively large set of 300 realisations is still
  not sufficient for a robust estimation of the full covariance of the
  hundreds of triangular configurations originally measured, we
  considered, in the first place, a likelihood analysis based on the
  bispectrum variance alone. In a second step, we rebinned the
  bispectrum measurements assuming a larger bin size for the
  wavenumber making up the triangle sides. This allows to reduce the
  overall number of triangle configurations to less than a hundred,
  allowing an estimate of their full covariance properties and the
  related likelihood analysis.

As in the similar analysis performed in the companion papers, we
assumed a model for the bispectrum and produced a data vector from the
evaluation of such model at some chosen fiducial value for the
parameters. This allowed us to focus our attention exclusively on the
errors recovered as a function of the different estimation of the
covariance matrix. Differently from the companion papers, the model
considered based on tree-level perturbation theory, only depends on
bias and shot-noise parameters, allowing a much easier evaluation of
the likelihood function. In particular, under these simplified
settings we can easily compute our results as a function of the
smallest scale, or maximum wavenumber $k_{max}$, included in the
analysis. More rigorous tests involving additional cosmological
parameters, a more accurate modelling of the redshift-space bispectrum
in the quasi-linear regime and a solid estimate of the full bispectrum
covariance matrix (and cross-correlation with the power spectrum) are
clearly well beyond the scope of this comparison project but will be
required in the near future for the proper exploitation of the galaxy
bispectrum as a relevant observable.

The parameter error comparison has shown that: 
\begin{enumerate}
\item
the error on the bias and on the shot-noise parameters are reproduced
within 10$\%$ by all the methods except Lognormal and \halogen in the
high-mass sample for $k_{max}>0.07$. This is evident as well in terms
of the combined error volume as defined in \eq{eq:vol}; for the second
sample \pin, and to a lesser extent \patchy, show an higher level of
disagreement compared with the other predictive methods;
\item
the Gaussian prediction tends to underestimate the error on some
parameters for large values of $k_{max}$;
\item
the two-parameters contour plots from the variance-based
  likelihood, for both mass samples and for different values of
  $k_{max}$ (not all shown in the figures), do not show any relevant
  difference among the methods in terms of parameter degeneracies;
  some differences in the recovered parameters degeneracies between
  the N-body and the \halogen, LogNormal, and, to a lesser extent
  \patchy results, are instead present in the case of the
  covariance-based likelihood.
 \end{enumerate}

To sum up, predictive methods, along with \patchy appear to be the
most accurate in reproducing the N-body results, but differences are
overall relatively small. Of course, due to the relatively small
number of N-body runs available, our likelihood tests have been
either limited to include the bispectrum variance or forced to
consider quite a larger $k$-bin, smoothing the shape-dependence of
the bispectrum and increasing the relevance of the off-diagonal
elements of its covariance matrix. For this reason, we included
additional tests employing 10,000 \pin realisations to compare, at
least for this particular method, the variance estimated from 300
realisation to the variance and the full covariance estimated from the
whole set. This has shown that
\begin{enumerate}
\item
the variance estimate is not particularly affected by the limited
number of 300 runs and essentially no difference is found on the
results for the parameters errors;
\item
the results in terms of the full covariance, instead, do provide
differences on the parameters errors but still within 10\%, although
they highlight a progressive underestimate of the errors based on the
variance alone beyond $k_{max}\simeq 0.15 \kMpc$, where a steady
deviation proportional to $k_{max}$ is observed.
 \end{enumerate}
Clearly, a more realistic investigation of the relevance of a
reliable estimate of the bispectrum covariance matrix requires a
proper model for the quasi-linear regime that we will leave for
future work. In addition, we should also expect that the relatively
small difference between the results obtained from the variance alone
and the full covariance will become more relevant once a realistic
window function is accounted for as beat-coupling/super-sample
covariance effects are expected to provide additional contributions
also to off-diagonal elements. Since such effects depend directly on
the non-Gaussian properties of the galaxy/halo distribution, we
consider the present work only as a first step toward a more complete
assessment of the correct recovery of non-Gaussianity by approximate
methods for mock catalogs.

From the analysis we have presented it appears that most of the
methods we considered are capable to reproduce the halo bispectrum,
its variance and the errors on bias parameters based on the variance
alone quite accurately. This is particularly true for predictive
methods such as \cola, \pin and \peak. Similar results are obtained
for \patchy, although the calibration in redshift space might lead to
some larger systematic for the real-space bispectrum that in turn
could have effects not investigated in this work (\eg finite-volume
effects). For what concern \halogen, we have already stressed that its
calibration is restricted to the two-point statistic so a lower
accuracy on the bispectrum might be somehow expected. Nevertheless is
worth to point out that the marginalized errors on the parameters in
redshift space, in particular for the first sample, are certainly
comparable with all the other methods except for Lognormal. This last
method, in fact, is the one that fares worst among those
considered. This is also not surprising since, as already mentioned,
the nonlinear transformation on the density field that provides a
qualitatively reasonable description of the nonlinear power spectrum,
while providing a non-Gaussian contribution, does not ensure that such
contribution, for instance in the case of the bispectrum, presents the
correct functional form and dependence on the triangular configuration
shape.

We notice finally how our tests on the bispectrum have highlighted
differences among the different methods that are less evident from the
similar analysis on two-point statistic performed in the companion
papers I and II. This illustrates how the bispectrum can be a useful
diagnostic for this type of comparisons, even when we are not directly
interested in the bispectrum as an observable. We expect that possible
direction of investigation along these lines will include correlators
of realistic galaxy distribution and, particularly for Fourier-space
statistics, finite-volume effects, in order to better assess the
interplay between non-Gaussianity, convolution with a window function
and realistic shot-noise contributions.

\section*{Acknowledgments}

We are grateful to the anonymous referee for a careful reading of the manuscript and, in particular, for suggesting the additional tests based on the full bispectrum covariance that, while not affecting the main outcomes of this work, significantly strengthen them.

M. Colavincenzo is supported by the {\em Departments of Excellence
  2018 - 2022} Grant awarded by the Italian Ministero dell'Istruzione,
dell'Universit\`a e della Ricerca (MIUR) (L. 232/2016), by the
research grant {\em The Anisotropic Dark Universe} Number CSTO161409,
funded under the program CSP-UNITO {\em Research for the Territory
  2016} by Compagnia di Sanpaolo and University of Torino; and the
research grant TAsP (Theoretical Astroparticle Physics) funded by the
Istituto Nazionale di Fisica Nucleare (INFN). P. Monaco and
E. Sefusatti acknowledge support from a FRA2015 grant from MIUR PRIN
2015 {\em Cosmology and Fundamental Physics: illuminating the Dark
  Universe with Euclid} and from Consorzio per la Fisica di Trieste;
they are part of the INFN InDark research group.

L. Blot acknowledges support from the Spanish Ministerio de Econom\'ia y Competitividad (MINECO) grant ESP2015-66861. M.Crocce acknowledges support from the Spanish Ram\'on y Cajal MICINN program. M.Crocce has been funded by AYA2015-71825. 

M. Lippich and A.G.S\'anchez acknowledge support from the Transregional Collaborative Research Centre TR33 {\em The Dark Universe} of the German Research Foundation (DFG). 

C. Dalla Vecchia acknowledges support from the MINECO through grants
AYA2013-46886, AYA2014-58308 and RYC-2015-18078. S. Avila acknowledges
support from the UK Space Agency through grant
ST/K00283X/1. A. Balaguera-Antol\'{i}nez acknowledges financial
support from MINECO under the Severo Ochoa program
SEV-2015-0548. M. Pellejero-Ibanez acknowledges support from MINECO
under the grand AYA2012-39702-C02-01. P. Fosalba acknowledges support
from MINECO through grant ESP2015-66861-C3-1-R and Generalitat de
Catalunya through grant 2017-SGR-885. A. Izard was supported in part
by Jet Propulsion Laboratory, California Institute of Technology,
under a contract with the National Aeronautics and Space
Administration. He was also supported in part by NASA ROSES
13-ATP13-0019, NASA ROSES 14-MIRO-PROs-0064, NASA ROSES 12-
EUCLID12-0004, and acknowledges support from the JAE program grant
from the Spanish National Science Council (CSIC). R. Bond, S. Codis
and G. Stein are supported by the Canadian Natural Sciences and
Engineering Research Council (NSERC). G. Yepes acknowledges financial
support from MINECO/FEDER (Spain) under research grant
AYA2015-63810-P.

The Minerva simulations have been performed and analysed on the Hydra and Euclid clusters at the Max Planck Computing and Data Facility (MPCDF) in Garching.

\pin mocks were run on the GALILEO cluster at CINECA thanks to an agreement with  the University of Trieste.

\cola simulations were run at the MareNostrum supercomputer - Barcelona Supercomputing Center (BSC-CNS, www.bsc.es), through the grant AECT-2016- 3-0015. 

\peak simulations were performed on the GPC supercomputer at the SciNet HPC Consortium. SciNet is funded by: the Canada Foundation for Innovation under the auspices of Compute Canada; the Government of Ontario; Ontario Research Fund - Research Excellence; and the University of Toronto.

Numerical computations with \halogen were done on the Sciama High Performance Compute (HPC) cluster which is supported by the ICG, SEPNet and the University of Portsmouth.

\patchy mocks have been computed in part at the MareNostrum supercomputer of the Barcelona Supercomputing Center thanks to a grant from the Red Espa\~nola de Supercomputaci\'on (RES), and in part at the Teide High-Performance Computing facilities provided by the Instituto Tecnol\'ogico y de Energ\'{\i}as Renovables (ITER, S.A.).  

This work, as the companion papers, has been conceived and developed as part of the join activity of the ``Galaxy Clustering'' and the ``Cosmological Simulations'' Science Working Groups of the Euclid survey consortium.

This paper and companion papers have benefited of discussions and the
stimulating environment of the Euclid Consortium, which is warmly
acknowledged.
\bibliographystyle{mnras}
%\bibliography{/Users/emi/Cosmo/Letteratura/bibliografie/cosmologia}
\bibliography{biblio}

\appendix

%%%%%%%%%%%%%%%%%%%%%%%%%%%%%%%%%%%%%%%%%%%%%%%%%%%%%%%%%%%%%%%%%
%%%%%%%%%%%%%%%%%%%%%%%%%%%%%%%%%%%%%%%%%%%%%%%%%%%%%%%%%%%%%%%%%
\section{Mass-cut vs Abundance matching}
\label{app:masscuts}

\begin{figure*}[t!]
\begin{center}
\includegraphics[width= 0.9\textwidth]{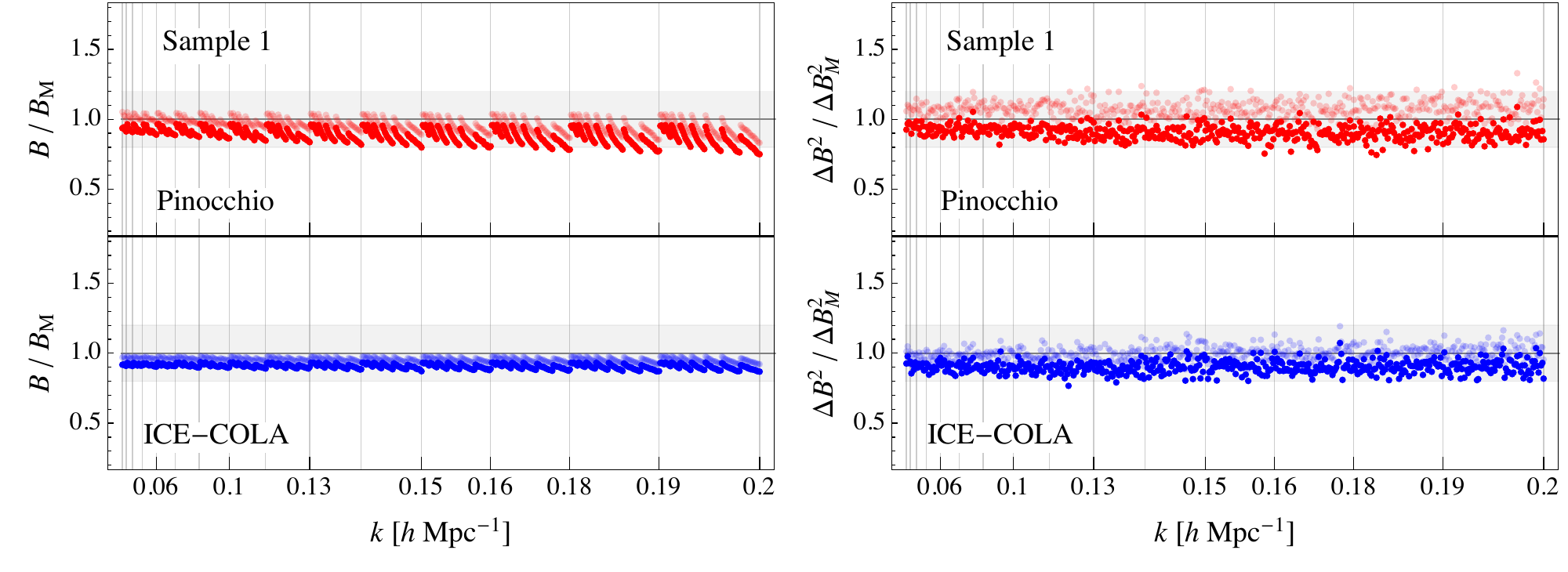}
\includegraphics[width= 0.9\textwidth]{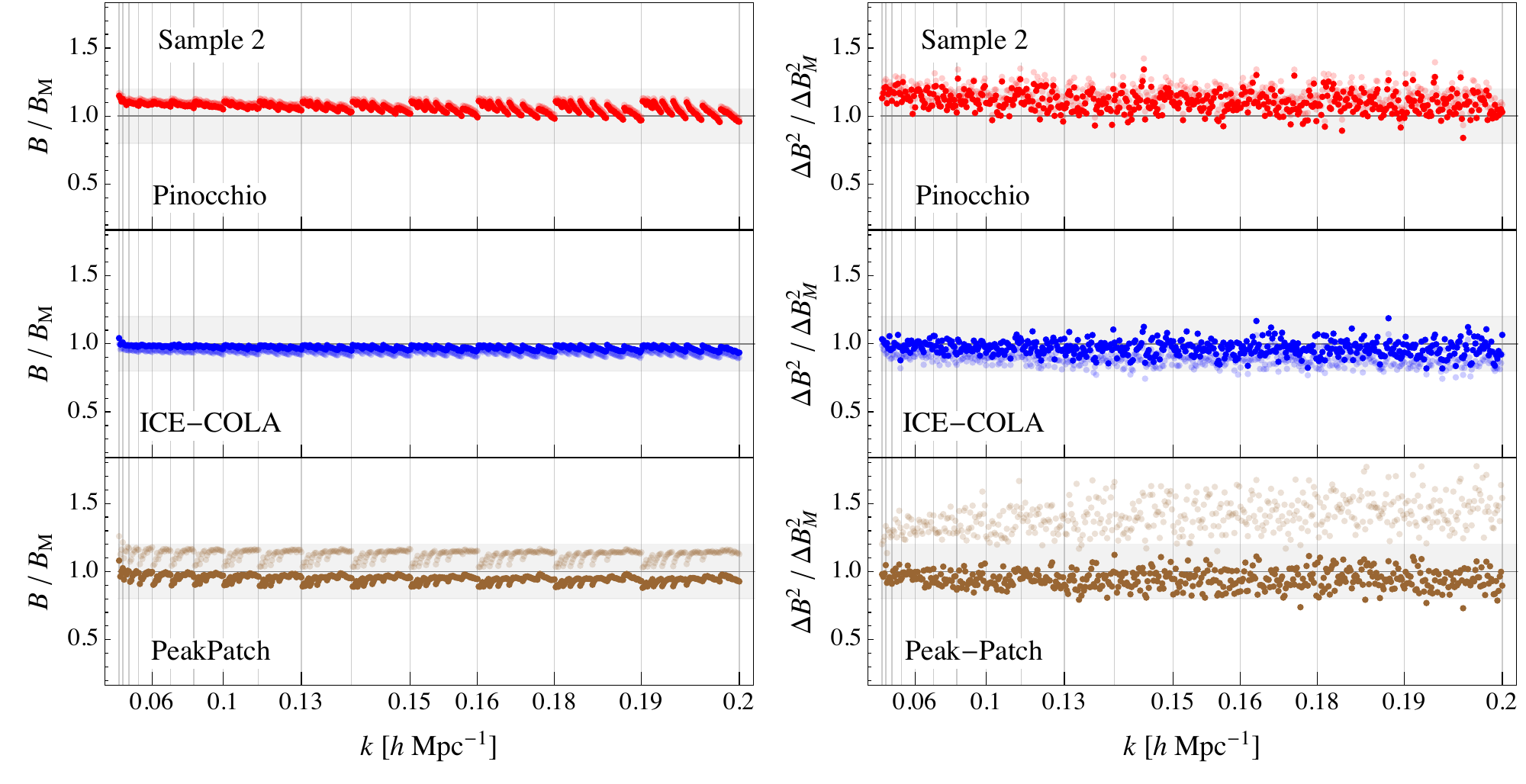}
\caption{\label{fig:bsMass} Bispectrum and its variance. Comparison of
  density matching (full color) to mass-cut (faded color), redshift
  space. }
\end{center}
\end{figure*}

We have seen how predictive methods perform better overall than
methods requiring calibration with a set of N-body
simulations. However, all our results did assume, including predictive
ones, that the halo density matches the one from the N-body catalogs
to mach the halo density from the N-body catalogs.
In this appendix we compare the results presented
so far and those obtained from \pin, \cola and \peak when
their predictions are taken out-of-the-box with no abundance matching. Since each method has a different definition of the mass, a constant mass cut will typically pick up different objects. This is especially true for \peak halos which are defined as spherical overdensities in Lagrangian space and are not meant to reproduce FoF masses.

Figure \ref{fig:bsMass} shows the ratio of the bispectrum (left
column) and its variance (right column) to the N-body results
(similarly to figures~\ref{fig:bsRSAllD1} and \ref{fig:bsRSAllD2})
in redshift space comparing the case of density matching (full color)
assumed so far to the case where the limiting mass is not changed
(faded color). Both mass samples are shown and we remind the reader
that the \peak catalogs are only available for Sample 2.

\begin{figure*}
\begin{center}
\includegraphics[width= 0.9\textwidth]{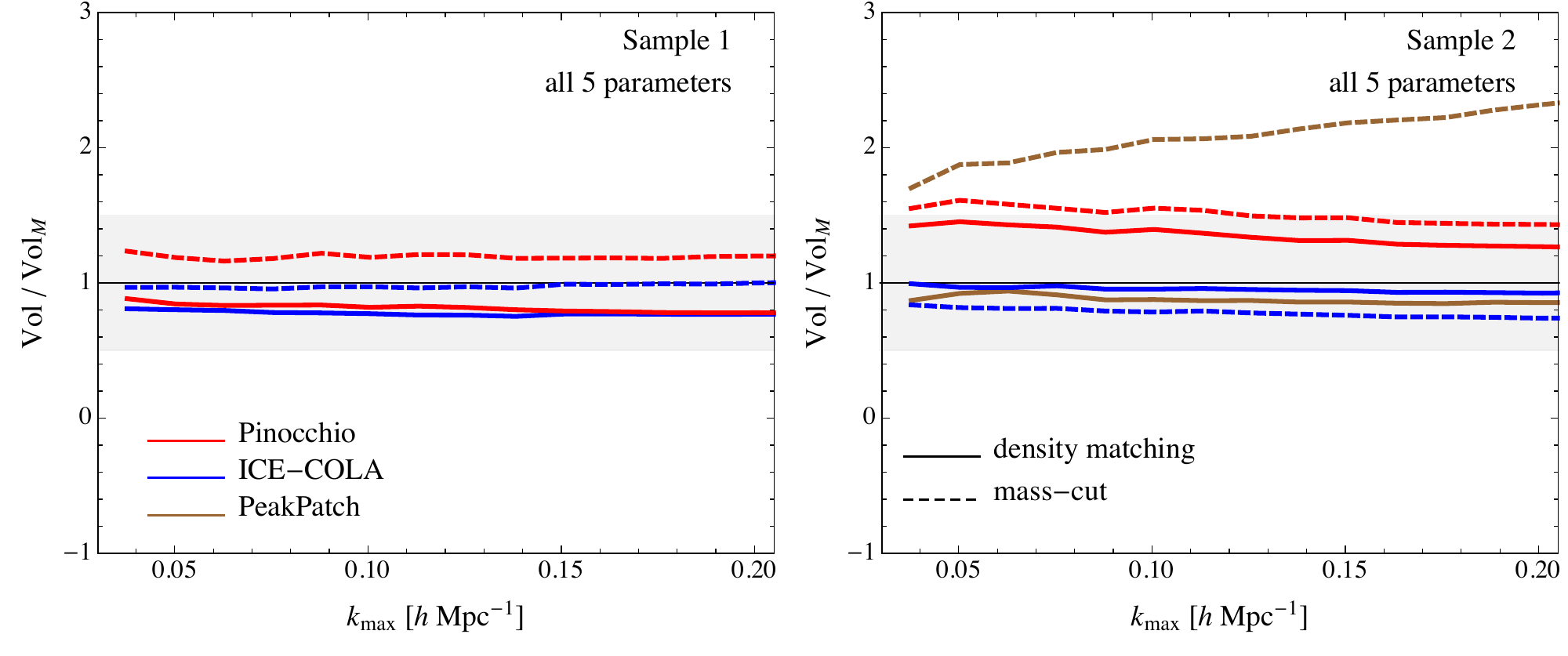}
\caption{\label{fig:bsMassErrors} Marginalized errors for the bias parameters
using the real bispectrum for the two samples (first and second column) compared with the 
error obtained using Minerva. Density cuts are displayed with solid lines while dashed lines represent mass cuts. The gray shaded area represent the 10$\%$ error on individual parameters, or 50\% on the 5-parameters error volume.}
\end{center}
\end{figure*}

For the bispectrum the difference between the density matching and the
mass-cut are lower than 10$\%$ for \pin and \cola for both the
samples, while \peak shows a larger difference, but always smaller
than 20$\%$, with density matching performing better as we can expect.
For the variance the differences appear to be larger. \cola and
\pin present, respectively, differences of the order of 15 to
25\% for the first sample but smaller in the second sample
case. \peak, on the other hand shows a difference of about 40\%
for Sample 2.

Finally, figure \ref{fig:bsMassErrors} shows the combined error volume relative to the N-Body results, as in figure
  \ref{fig:margErrRSD1D2}, for the two samples, comparing density matching (continuous lines) to the case of direct mass-cut (dashed
  lines). Using the measurements from the mass-cut case, for both samples, we recover larger errors, as can be expected from the variance comparison, with differences of the order of 10\% on the individual parameter error (50\% on the 5-parameter volume shown in the figure) for \pin. An even larger difference is found for \peak, while discrepancies for \cola are within 5\% for both samples.  

\end{document}